\documentclass[aps,prb,twocolumn,superscriptaddress,10pt,article,nofootinbib,longbibliography]{revtex4-2}
\usepackage{graphicx,amsfonts,amssymb,amsmath,hyperref,hypcap,enumerate,bm,color,float}
\usepackage{slashed}
\usepackage{comment}
\newcommand{\beq}{\begin{equation}}
\newcommand{\eneq}{\end{equation}}

\def\qq{\mathbf{q}}
\def\kk{\mathbf{k}}
\def\KK{\mathbf{K}}
\def\DKK{\Delta\mathbf{K}}
\def\pp{\mathbf{p}}
\def\RR{\mathbf{R}}
\def\tt{\mathbf{t}}
\def\rr{\mathbf{r}}
\def\GG{\mathbf{G}}
\def\QQ{\mathbf{Q}}
\def\aa{\mathbf{a}}
\def\bb{\mathbf{b}}
\def\uu{\mathbf{u}}
\def\AA{\mathbf{A}}
\def\ff{\mathbf{f}}
\def\ll{\mathbf{l}}

\def\KK{\mathbf{K}}
\def\qq{\mathbf{q}}
\def\pp{\mathbf{p}}
\def\pp{\mathbf{p}}
\def\GG{\mathbf{G}}
\def\QQ{\mathbf{Q}}
\def\RR{\mathbf{R}}
\def\tt{\mathbf{t}}
\def\dd{\mathbf{d}}
\def\aa{\mathbf{a}}
\def\bb{\mathbf{b}}
\def\ee{\epsilon}
\def\CC{\mathcal{P}}
\def\UU{\mathbf{U}}

\def\BZ{{\rm BZ}}
\def\mS{{\mathcal{S}}}

\def\spin{{\varsigma}}

\def\hH{{ \hat{H} }}
\def\hrho{ \hat{\rho} }
\def\hg{\hat{g}}
\def\hS{\hat{S}}

\def\mG{{\mathcal{G}}}

\def\UC{{\hat{\Theta}}}
\def\UF{{\hat{\Sigma}}}

\def\mK{{\mathcal{K}}}
\def\pr{\prime}
\def\mJ{{\mathcal{J}}}
\def\mK{{\mathcal{K}}}

\def\ie{{\it i.e.},\ }
\def\eg{{\it e.g.}\ }
\def\ea{{\it et al.}}

\begin{document}

\title{Localized interfacial Phonon Modes at the Electronic Axion Domain Wall}

\author{Abhinava Chatterjee$^{1}$, Mourad Oudich$^{2}$, Yun Jing$^{2}$, and Chao-Xing Liu$^{*}$}
\affiliation{%
   Department of Physics, The Pennsylvania State University, University Park, Pennsylvania 16802, USA \\
  $^{2}$ Graduate Program in Acoustics, The Pennsylvania State University, University Park, Pennsylvania 16802, USA
}

\begin{abstract} 
The most salient feature of electronic topological states of matter is the existence of exotic electronic modes localized at the surface or interface of a sample. In this work, in an electronic topological system, we demonstrate the existence of localized phonon modes at the domain wall between topologically trivial and non-trivial regions, in addition to the localized interfacial electronic states. In particular, we consider a theoretical model for the Dirac semimetal with a gap opened by external strains and study the phonon dynamics, which couples to electronic degrees of freedom via strong electron-phonon interaction. By treating the phonon modes as a pseudo-gauge field, we find that the axion type of terms for phonon dynamics can emerge in gapped Dirac semimetal model and lead to interfacial phonon modes localized at the domain wall between trivial and non-trivial regimes that possess the axion parameters 0 and $\pi$, respectively. We also discuss the physical properties and possible experimental probe of such interfacial phonon modes. 
\end{abstract}

\date{\today}

\maketitle

{\it Introduction - }
At the surface of a topological electronic material or at the domain wall between two topologically distinct regions, surface or interfacial electronic modes can emerge and exhibit exotic physical phenomena \cite{qi2011topological,hasan2010colloquium,yan2012topological,wieder2022topological}. The existence of these surface/interfacial electronic modes is a direct consequence of non-trivial topology of the bulk electronic band structure in topological materials, a general feature summarized as the bulk–boundary correspondence \cite{ryu2002topological,schnyder2008classification,mong2011edge,ryu2010topological,tanaka2011symmetry,kitaev2009periodic,hatsugai2009bulk}. The connection between bulk topological states and the corresponding boundary modes is not limited to electronic states, but has also been generalized to other physical systems, including photonics \cite{lu2014topological,ozawa2019topological}, phononics \cite{liu2020topological,chen2018chiral,chen2021topological}, magnonics \cite{zhuo2023topological,mcclarty2022topological}, electric circuits \cite{bergholtz2021exceptional,gilbert2021topological}, and mechanical systems \cite{zheng2022progress,ma2019topological,mao2018maxwell}, in which a variety of boundary modes have been identified and classified, and many of them have been experimentally observed. 

Our current understanding and classification of topological states and boundary modes (e.g. electrons, phonons, magnons, {\it etc},) is mainly limited to systems with one type of quasi-particles. As the interaction between different quasi-particles, e.g. electron-phonon interaction, generally exists in realistic materials \cite{thalmeier2011surface,giraud2011electron,giraud2012electron,huang2012surface,parente2013electron,garate2013phonon}, one may ask if topological states and the boundary modes of different quasi-particles are connected to each other. Topological polariton \cite{karzig2015topological,liu2020generation,klembt2018exciton,hu2020topological,li2021experimental,kartashov2019two} provides such an example. 
Here we wonder, in a system with strong electron-phonon interaction, if a boundary phonon mode can exist for a topologically non-trivial electronic system. 

In this work, we give an affirmative answer to this question and show that an interfacial phonon mode can exist at the domain wall between two topologically distinct regions with different electronic axion parameters. Particularly, we consider the Dirac semimetal model with a small gap controlled by external static strain serving as effective axion fields. Acoustic phonon dynamics is strongly influenced by the axion term through electron-acoustic-phonon (or equivalently electron-strain) interaction, which gives rise to interfacial phonon modes localized at the domain wall between topologically trivial and non-trivial regions of electronic states. 

\begin{figure*}
\includegraphics[width=\textwidth]{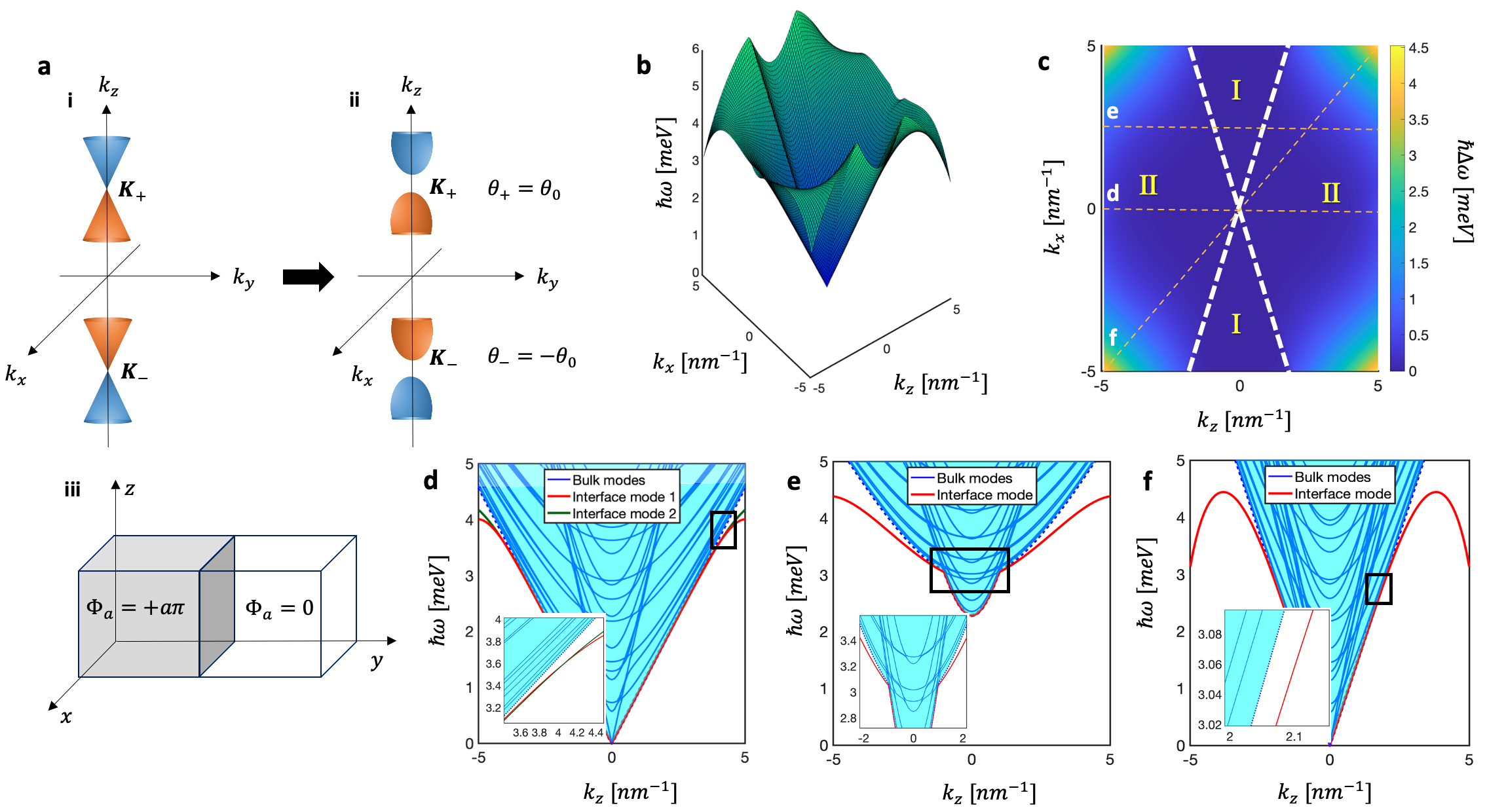} 
\caption{(a) Setup. (i) The spectrum for $Na_3Bi$ is gapless at the $K_\pm$ points. (ii) The external strain leads to a gap in the spectrum where $\theta_0 = \pi/2$ by choosing the static strain $u_{xy}^0 \rightarrow 0$ and $u_{xx}^0 - u_{yy}^0 \neq 0$. (iii) The variation of the static axion angle along the $y$ direction which induces a domain wall at the interface. (b) Interface mode dispersion $\hbar \omega$ as a function of in-plane momenta $(k_x, k_z)$. (c) Energy difference between the lowest bulk mode and the next lowest mode $\hbar \Delta  \omega = \hbar (\omega_2 - \omega_1)$. The thick white dashed lines separate Region I where the interface mode merges with the bulk modes, and Region II where the interfacial mode is below the bulk modes in energy. The thin dashed orange lines show cuts in the $k_xk_z$ plane for which we have plotted the energy dispersion in (d), (e), and (f). (d) Band dispersion at $k_x=0$ where two interface modes (marked in red and green lines) exist. The bulk modes are highlighted in blue and shaded with a cyan region. Inset: the band crossing between the two interface modes. (e) Band dispersion at $k_x=2.5\, \text{nm}^{-1}$ where one interface mode (red line) merges into the bulk modes for $k_z \sim [-1,1]\, \text{nm}^{-1}$ (Inset),consistent with Region I of  (c). (f) Band dispersion at $k_x=k_z$ where one interface mode (red) exists. Inset: the interface mode's velocity is than that of the the bulk modes. The lowest bulk mode is shown with the dotted blue line in (d)-(f). We use $r=.8 $; $s=.713 $; $t=.9 $ in units of $10^{-12} m^4/s^2$ and $a=6.75, b=3.17, c=1.01, d=1.93, f=11.02$ in units of $10^6\, \text{m}^2/\text{s}^2$.}
\label{fig:fig1}
\end{figure*}

{\it Electron-phonon Interaction and Valley Axion Field in Gapped Dirac Semimetals - } 
We consider three-dimensional (3D) Dirac semimetal, e.g. Na$_3$Bi \cite{liu2014discovery,xiong2015evidence}, with electron-phonon coupling, described by the Hamiltonian $H=H_e+H_{e-ph}$. $H_e$ is the model Hamiltonian for 3D Dirac semimetals \cite{wang2012dirac,wang2013three}, and its detailed form is described in Sec.S1 B of Supplementary Materials (SM) \cite{materialparameters}. 
The crystal symmetry of $H_e$ is described by 
the $D_{6h}$ group that can be generated by six-fold rotation $\hat{C}_{6z}$, inversion $\hat{I}$ and two-fold rotation $\hat{C}_{2x}$. The conduction and valence bands of $H_e$ have the crossings located at the momenta $\KK_{a}=(0,0,a k_c)$ with $a=\pm$ and $k_c$ depending on material parameters. Around these crossing points, the effective Hamiltonians behave as 3D massless Dirac fermions with nodes protected by $\hat{C}_{6z}$, as schematically depicted in Fig. \ref{fig:fig1}(a)(i). The index $a=\pm$ in $\KK_{a}$ is regarded as two "valleys" that are related by time reversal $\hat{T}$. For electron-phonon interaction, we consider acoustic phonons that can be described by the strain tensor $u_{ij} = \frac{1}{2}\left( \partial_i u_j + \partial_j u_i \right)$ ($i,j=x,y,z$) with $\uu$ labelling the displacement field \cite{landau1986theory}. Based on the $D_{6h}$ group for Na$_3$Bi, we can construct all the terms up to the linear orders in both the momentum $\kk$ and strain tensor $u_{ij}$ (See Sec.S1 B of SM) \cite{materialparameters}.

We next project the total Hamiltonian $H$ %$H = H_e + H_{e-ph}$ 
into the subspace spanned by the massless Dirac fermions at $\KK_a$, and the corresponding effective Hamiltonian is written as $H_{eff} = \sum_{a=\pm} H_{eff,a}$ with $H_{eff,a}$ given by 

\begin{equation}\label{Ha}
\begin{split}
    H_{eff,a} &= A_0^{pse} \mathbb{I} + A_0 {\bf \pi}_x  \Gamma_3 - A_0 {\bf \pi}_y \Gamma_4  - 2 M_1 k_c {\bf \pi}_z \Gamma_5 \\ & + |m| \left( \cos\Phi_a \Gamma_1 - \sin\Phi_a \Gamma_2 \right),
\end{split}
\end{equation}
%|m| e^{i \Phi_a \sigma_3 \otimes I}
where $\bm{\pi} = \kk - \AA^{pse}_a$ with the pseudo-gauge field \cite{de2013gauge,pikulin2016chiral,ilan2020pseudo,yu2021pseudo} $A_0^{pse} = \Tilde{C}(\{u_{ij}\})$ and $\textbf{A}^{pse}_a = a \left[ - A_4 k_c u_{xz} , A_4 k_c u_{yz} , \Tilde{M}(\{u_{ij}\})\right]$ ($a=\pm$). Here $\Tilde{C}(\{u_{ij}\}) = C_3 u_{zz} + C_4 (u_{xx} + u_{yy})$, $\Tilde{M}(\{u_{ij}\}) = M_3 u_{zz} + M_4 (u_{xx} + u_{yy})$, $|m| = D k_c \sqrt{(2u_{xy})^2 + (u_{xx}-u_{yy})^2}$, 
$\Phi_a = a \theta $, 
$\cot\theta = \frac{u_{xx}-u_{yy}}{2 u_{xy}} $, with $C_{3,4},M_{3,4},A_{0,4}, D$ as material parameters \cite{materialparameters}. $|m| \left( \cos\Phi_a \Gamma_1 - \sin\Phi_a \Gamma_2 \right)$ is a complex valley dependent mass term due to strain that can gap out the Dirac points, as depicted in Fig. \ref{fig:fig1}(a) (ii). We consider a static strain $u^{(0)}_{ij}$ and a dynamical strain field $\delta u_{ij}$,  $u_{ij} = u^{(0)}_{ij} + \delta u_{ij}$, in which the static strain is chosen only to possess non-zero $u^{(0)}_{xx}-u^{(0)}_{yy}$ and $u^{(0)}_{xy}$ while all other components are zero. Consequently, $u^{(0)}_{ij}$ can produce the $\Phi_a$ field but not the pseudo-gauge field $\mathcal{A}^{pse} =(A_0^{pse},\textbf{A}^{pse}_a)$. 
We consider acoustic phonons that create a dynamical strain field $\delta u_{ij}$, leading to the pseudo-gauge field $\mathcal{A}^{pse}$. We notice that $A_0^{pse} $ has the same sign for both valleys, while $\AA^{pse}_a$ flips its sign between two valleys, as required by time reversal $\hat{T}$.  

 \begin{figure*}
 \includegraphics[width=\textwidth]{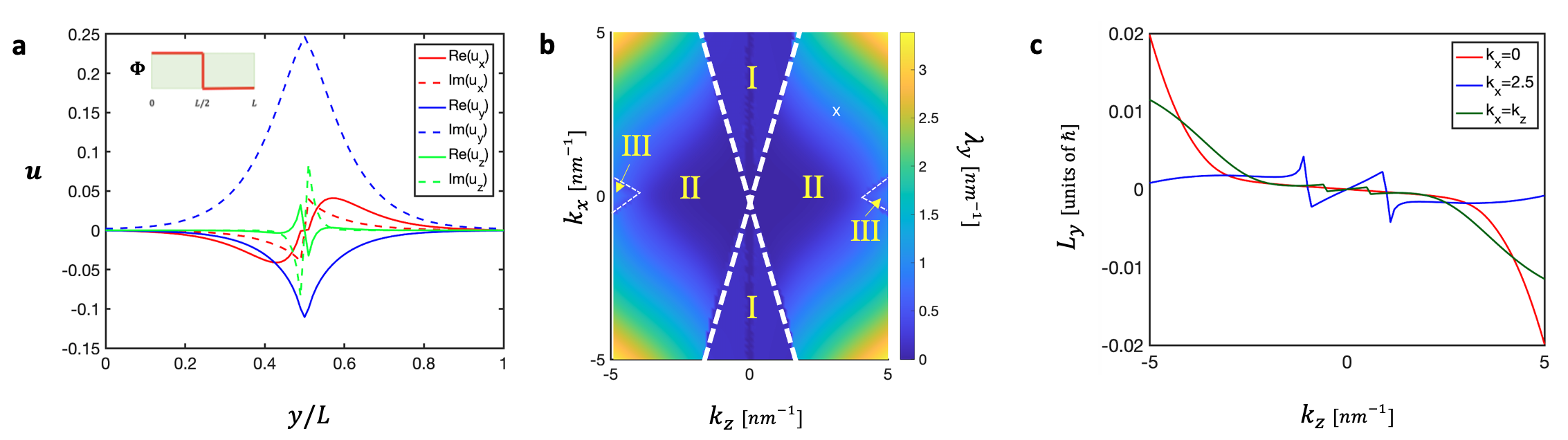}
 \caption{(a) The spatial distribution of the real and imaginary parts of the displacement field of the interface phonon mode at point X where $(k_x,k_z) = (2.5,3) nm^{-1}$. The $u_y$ component (both real and imaginary) is even across the axion domain wall whereas the $x$ and $z$ components are odd. (b) The inverse decay length ($\Lambda$) of the dominant $u_y$ component of the displacement as a function of $(k_x,k_z)$ (c) The out of plane ($y$ direction) phonon angular momentum of the interface mode at $k_x=0$ (red) , $k_x=2.5 nm^{-1}$ (blue) and $k_x=k_z$ (dark green). We use $r=.8 $; $s=.713 $; $t=.9 $ in units of $10^{-12} m^4/s^2$ and $a=6.75, b=3.17,c=1.01,d=1.93,f=11.02$ in units of $10^6 m^2/s^2$. 
}
 \label{fig:fig2}
 \end{figure*}

Next we assume the Fermi energy is within the energy gap of the effective Hamiltonian Eq.\ref{Ha}, so that we can integrate out the Dirac fermions  and the resulting effective action of the pseudo-gauge field $\mathcal{A}^{pse}$ reads \cite{yu2021dynamical}
\begin{eqnarray} \label{eq:axion_term}
& S_{ax} = \sum_{a=\pm} S_{eff,a}, \nonumber \\
& S_{eff,a} = \frac{1}{32 \pi^2} \int dt d^3 r \Phi_a \epsilon^{\mu \nu \rho \delta} F_{\mu \nu,a} F_{\rho \delta,a},
\end{eqnarray}

where $F_{\mu \nu, a} = \partial_{\mu} \mathcal{A}_{\nu,a}^{pse} - \partial_{\nu} \mathcal{A}_{\mu,a}^{pse}$. $S_{eff,a}$, $\mu, \nu = 0,1,2,3$, has a similar form as the axion electrodynamics \cite{wilczek1987two,sekine2021axion,nenno2020axion}, but it is for pseudo-gauge field $\mathcal{A}^{pse}$ connected to the strain field. Thus, $S_{eff,a}$ is expected to influence phonon dynamics. %, instead of electromagnetic fields. Second, as $T$ is preserved in our original Hamiltonian Eqs.(\ref{eq:HNa3Bi}) and (\ref{eq:Heph}), 
Due to time reversal $\Hat{T}$, the axion field $\Phi_a$ has opposite signs for two valleys, and thus we dub $\Phi_a$ as the "valley axion field".
The derivation for the explicit form of the total effective action $S_{ax}$ can be found in Sec.S1 C of SM \cite{materialparameters}. 
The effective action is invariant under $\Hat{C}_{6z}$, $\Hat{C}_{2x}$, $\Hat{T}$ and $\Hat{I}$. We can re-write the effective action in a compact form
\begin{eqnarray} \label{eq:Sax_compact}
    S_{ax} = - \int dt d^3r \left( \nabla \theta \cdot \textbf{h} \right), 
\end{eqnarray}
where $h_x = - \Tilde{C} \left( a A_0 A_4 M_0 \partial_z u_{yz} + A_0^2 \partial_y \Tilde{M} \right)$, $h_y = \Tilde{C} \left( A_0 A_4 M_0 \partial_z u_{xz} + A_0^2 \partial_x \Tilde{M} \right)$ and $h_z = A_0 A_4 M_0 \Tilde{C} \left(   \partial_x u_{yz} -  \partial_y u_{xz} \right)$. $S_{ax}$ vanishes when $\theta$ is a constant and is non-zero at the $\theta$ domain wall.

{\it Interficial Phonon Modes at Axion Domain Wall- } From the functional derivative $\delta (S_0+S_{ax})/\delta \textbf{u}=0$, where $S_0$ is the bulk elastic action for the $D_{6h}$ group with 5 independent elastic modulii $a,b,c,d,f$ \cite{de2015charting,dong2019electronic} and $S_{ax}$ is the axion term in Eq. (\ref{eq:axion_term}), one can derive the equation of motion for the phonon displacement field $\textbf{u}$ as 
\begin{eqnarray} \label{eq:EOM_ph}
    H_{ph} \textbf{u} = \rho \omega^2 \textbf{u}
\end{eqnarray}, where $H_{ph}$ is the effective phonon Hamiltonian given by $H_{ph} = H_{bulk} + H_{ax}$ 
(See Sec. S1 F and Sec. S2 A of SM \cite{materialparameters} for the explicit form of $H_{ph}$). 
Here we assume the translation symmetry along the x and z directions, so that the in-plane momentum $\textbf{k}_{\parallel} = (k_x,k_z)$ a good quantum number. Along the y direction, we consider a domain wall configuration for the axion field, $\Phi_a = a \theta(y) = a \pi \Theta(L/2-y)$ , where $\Theta(y)$ is the step function, $a=\pm$ and $L$ is the system size as depicted in In Fig. \ref{fig:fig1}(a) (iii). This $\Phi$ configuration can be achieved by applying a static strain $u^{(0)}_{xx} - u^{(0)}_{yy}>0$ for $y<L/2$ and $u^{(0)}_{xx} - u^{(0)}_{yy}<0$ for $y>L/2$ in the limit $u^{(0)}_{xy} \rightarrow 0$. Since $\Delta \Phi = \pi$, the interface is an electronic axion domain wall. 
By choosing the wave-function ansatz $\textbf{u} (\rr,t) = \textbf{f}(y) e^{i \textbf{k}_{\parallel}\cdot \textbf{r}_{\parallel}}$ with $\textbf{r}_{\parallel}=(x,z)$ and $\kk_\parallel=(k_x,k_z)$, the form of the Hamiltonian  $H_{ax}$ is given by (See Sec. S2 A of SM \cite{materialparameters} )
\begin{widetext}

\begin{equation} \label{Hax}
    H_{ax}  = \begin{pmatrix} 
     0  &  \frac{s}{2} k_z^2 \{ \partial_y, \frac{\partial \theta}{\partial y}\} & i \frac{\partial \theta}{\partial y} \alpha(k_x,k_z) \\ 
       -\frac{s}{2} k_z^2 \{\partial_y,\frac{\partial \theta}{\partial y}\} & 0 & -\frac{1}{2} (s + t) k_x k_z  \{\partial_y,\frac{\partial \theta}{\partial y}\} \\
     -i \frac{\partial \theta}{\partial y} \alpha(k_x,k_z) & +\frac{1}{2} (s + t) k_x k_z \{\partial_y,\frac{\partial \theta}{\partial y}\} & 0
    \end{pmatrix}.
\end{equation} 
\end{widetext}
where $\alpha(k_x,k_z) = r k_z^3 - (s + t) k_x^2 k_z $, $ \frac{\theta_0}{2\pi^2 Z} \frac{J}{2}C_3 \equiv r$, $ \frac{\theta_0}{2\pi^2 Z} \frac{J}{2}C_4 \equiv s$ and $\frac{\theta_0}{2\pi^2 Z} A_0^2 (C_3 M_4 - C_4 M_3) \equiv t$ with $J = -2 A_0 A_4 M_0$ and $Z = 2 A_0^2 M_1 k_c$. $H_{ax}$ contains $\frac{\partial \theta}{\partial y}$, which is only non-zero at the axion domain wall.

We numerically solve the full Hamiltonian $H_{ph}$ (Sec. S3 in SM \cite{materialparameters}). 
Fig. \ref{fig:fig1}(b) illustrates the anisotropic dispersion of the lowest frequency phonon mode in the $\kk_{\parallel}$ plane. Fig. \ref{fig:fig1}(d) and (e) show the phonon dispersion along the $k_z$ direction for $k_x=0$ and $k_x=2.5 nm^{-1}$, respectively, while Fig. \ref{fig:fig1}(f) is for phonon dispersion along the $k_x=k_z$ direction. The cyan shaded areas in Fig. \ref{fig:fig1}(d)-(f) represent the dispersion of bulk phonons that can be obtained from the eigen-frequency of $H_{bulk}$. In the insets of Fig. \ref{fig:fig1}(d)-(f), additional phonon modes emerge with their frequency below the bulk phonon frequency. We find one mode (red) in Fig. \ref{fig:fig1}(e) and (f), and two modes (red and green) in Fig. \ref{fig:fig1}(d). Fig. \ref{fig:fig1}(c) depicts the frequency difference between the lowest bulk phonon frequency and the lowest phonon modes in the domain wall configuration in the $\kk_\parallel$ space, which can be divided into two regions. This frequency difference is zero in Region I, while it becomes positive in Region II, implying the existence of the modes with their frequencies below bulk phonon frequency. To further demonstrate their origin from the axion domain wall, the wave function $\ff(y)$ for the $y$-directional distribution of the displacement field at $X=(k_x,k_z)= (2.5,3) nm^{-1}$ is depicted in Fig. \ref{fig:fig2}(a). All components of displacement field $\uu$ decay exponentially away form the the axion domain wall located at $y=L/2$, thus unambiguously demonstrating the interfacial nature of these phonon modes. The interfacial phonon wave functions $\ff(y)$ for different momenta are further discussed in Fig. S2 of SM \cite{materialparameters}. We can use the exponential function $\sim e^{-\lambda_y |y-L/2| }$ to fit to the solution $f_y(y)$ that exhibits the largest amplitude for different $\kk_\parallel$, and the $\lambda_y$ parameter that represents the inverse of localization length is shown as a function of $\kk_\parallel$ in Fig. \ref{fig:fig2}(b). $\lambda_y=0$ in Region I and nonzero $\lambda_y$ appears in Region II and III, which means the interfacial modes only exist in Regions II and III, consistent with Fig. \ref{fig:fig1}(c). Region II has one interfacial mode while Region III contains two interfacial modes. More information of these two interfacial modes can be found in Fig. S5 of SM \cite{materialparameters}.
We also find with increasing $|\kk_\parallel|$, $\lambda_y$ increases, so that the localization length decreases. Thus, the interfacial phonon modes become more localized for a large momentum. The existence of the interfacial phonon modes %and their angular momentum properties 
can be understood by analytically solving the eigen-equation (\ref{eq:EOM_ph}) with the domain wall configuration under certain approximations. Two analytically solvable cases are considered in Sec. S2 B,C \cite{materialparameters}, and \textit{Case I} shows the presence of interfacial modes with non-zero angular momentum, 
%such as the red bands in Fig. \ref{fig:fig1}(d)-(f), 
whereas \textit{Case II} shows the presence of two interface modes, similar to the case in Fig. \ref{fig:fig1}(d). 
Below we will only focus on \textit{Case I} while \textit{Case II} is discussed in Sec.S2.C of SM \cite{materialparameters}. 

For \textit{Case I}, we choose $r \neq 0$ and $s = t =0$, such that only $\alpha = r k_z^3 \neq 0$ and take the isotropic approximation for $H_{bulk}$ with $a = c_l^2 - c_t^2, b = d = c_t^2, c = c_l^2 - 2c_t^2, f = c_l^2$. We further assume $c_t = c_l = c_0$ such that
\begin{equation} \label{isoxzmatrix}
H_{ph} = \begin{pmatrix}
    c_0^2 k^2 - c_0^2 \partial_y^2 & 0 & - i \alpha(k_x,k_z) \delta(y) \\
    0  & c_0^2 k^2 - c_0^2 \partial_y^2  & 0 \\
    i \alpha(k_x,k_z) \delta(y) & 0 &  c_0^2 k^2 - c_0^2 \partial_y^2 
    \end{pmatrix}.
\end{equation}

Due to the $\delta(y)$-function, we look for exponentially localized phonon modes around $y=0$ with the ansatz $\textbf{u} = \sum_{\tau=1}^{3} A_\tau e^{-\lambda_\tau y} \textbf{u}_0^\tau(y,\lambda_\tau)$ for $y>0$ and $\textbf{u} = \sum_{\tau=1}^{3} B_\tau e^{\xi_\tau y} \textbf{v}_0^\tau(y,\lambda_\tau)$ for $y<0$, where $\lambda_\tau$ and $\xi_\tau$ are the inverse localization lengths of mode $\tau$ for $y>0$ and $y<0$ respectively. $\uu_0$ and ${\bf v}_0$ are the eigenmodes for $y>0$ and $y<0$ respectively. 
By matching the wave functions at the boundary at $y=0$, we obtain 6 linear equations for the variables $A_{1,2,3}$ and $B_{1,2,3}$. By solving for the characteristic equation, we find bulk modes with $\omega = c_0 k$ ($k=\sqrt{k_x^2+k_z^2}$) and an interfacial mode with $\omega = \left[(c_0 k)^2 - \left( \alpha(k_x,k_z)/c_0 \right)^2 \right]^{1/2}$. 
The eigen-vector of the interfacial phonon mode is $\textbf{u} = N e^{-\lambda |y|} ( \mathrm{sgn}(\alpha),0,-i)^T$ with the normalization factor $N$ and the inverse localization length $\lambda = |\alpha(k_x,k_z)|/c_0^2$. 
This interfacial phonon mode is circularly polarized and the corresponding phonon angular momentum, defined by $l_i = \hbar \textbf{u}_0^{\dagger} M_i \textbf{u}_0$ where $i=x,y,z$ and $(M_i)_{jk} = (-i) \epsilon_{ijk}$ \cite{zhang2014angular,hamada2018phonon}, has non-zero y-component, $l_y = \mathrm{sgn}(\alpha) \hbar = \mathrm{sgn}{k_z} \hbar$. The quantized angular momentum $l_y$ is a consequence of the isotropic approximation. For a more realistic situation, we numerically evaluate the angular momentum of the interfacial phonon modes, as depicted in Fig. \ref{fig:fig2}(c), in which $l_y$ strongly depends on $k_z$ and increases rapidly for a large $k_z$. $l_y$ changes its sign for opposite momentum $k_z$ and thus reveals a helical nature due to time reversal $\hat{T}$ \cite{hu2021phonon,liu2022probing}. More information of phonon angular momentum is provided in Fig. S3 and S4, and discussed in Sec. S5 of SM \cite{materialparameters}. This phonon helicity can be probed via temperature gradient induced phonon angular momentum \cite{hamada2018phonon,hu2021phonon,liu2022probing}, and our numerical and symmetry analysis suggested the z-directional temperature gradient can induce the y-directional angular momentum $l_y$ via the axion term, and this response is absent if there is no axion term, as discussed in Sec.S5.B and C of SM (Fig. S9b) \cite{materialparameters}. 
With increasing $|\kk_\parallel|$, the dispersion of the interfacial phonon mode generally exhibits a maximum at a certain momentum and then its frequency drops down. With further increasing $\kk_\parallel$, the phonon frequency can drop to zero and then become imaginary, implying a lattice instability. Nevertheless, our effective action of acoustic phonon dynamics is only valid for the momentum within the first Brillouin zone and breaks down closer to the Brillouin zone boundary (roughly at $\sim 10 nm^{-1}$), so we restrict our in-plane momenta to be $5 nm^{-1}$. Within this range of momenta and based on the current material parameters, the axion term $H_{ax}$ is not strong enough for a lattice instability.
It is well-known that a surface acoustic wave can exist at the surface of an elastic material (described by the phonon Hamiltonian $H_{bulk}$ with a stress-free boundary condition), even in the absence of the valley axion term \cite{landau1986theory}. The presence of axion term will also strongly affect the surface acoustic wave (See Sec. S4 of SM \cite{materialparameters}). % \ref{secsurf} \cite{materialparameters}).   

{\it Conclusion - }
In this work, we demonstrate the existence of the interfacial phonon modes localized at the domain wall between two regions with different valley axion parameters, which originates from the electron-phonon interaction in the gapped Dirac semimetal model. Although our model is derived for Na$_3$Bi, this mechanism can be generalized to other Dirac materials. The key requirement is that the low-energy Dirac Hamiltonian should be located at generic momenta that are not time-reversal-invariant, as the complex mass and the pseudo-gauge field terms in Eq. (\ref{Ha}) generally break $\hat{T}$. Given this requirement, the existing Dirac materials, $Na_3Bi$ and $Cd_3As_2$ \cite{wang2012dirac,wang2013three} may host interfacial phonon modes from our mechanism. We notice that the interfacial phonon modes have been previously observed in epitaxial Si-Ge interface via a combination of Raman spectroscopy and high-energy-
resolution electron energy-loss spectroscopy (EELS) in a scanning transmission electron microscope\cite{cheng2021experimental}, and in epitaxial cubic boron nitride/diamond heterointerface using 4D EELS \cite{qi2021measuring}, and similar experimental probes may be implemented to detect the interfacial phonon modes. The unique phonon angular momentum distribution in the momentum space\cite{hamada2018phonon,hu2021phonon,liu2022probing} may help to distinguish this mechanism of interfacial phonon modes from other origins \cite{cheng2021experimental,qi2021measuring}. Our mechanism can also be applied to magnetic topological materials, e.g. MnBi$_2$Te$_4$, even though the Dirac Hamiltonian is located at a high-symmetry momenta \cite{otrokov2019prediction,zhang2019topological,li2019intrinsic} but $\hat{T}$ is already broken by its intrinsic magnetism. One may also expect interfacial magnon mode in these magnetic topological materials due to electron-magnon interaction.  

{\it Acknowledgement} - We acknowledge helpful discussions with Z. Bi and N. Kalyanapuram. A.C. and C.-X.L. acknowledge support from NSF grant via the grant number DMR-2241327.

\bibliographystyle{apsrev4-2}

\bibliography{main}

\onecolumngrid

\renewcommand{\thefigure}{S\arabic{figure}}
%\addtolength{\textwidth}{3pt}
%\addtolength{\textheight}{0pt}
\renewcommand{\thesection}{S\arabic{section}}
\renewcommand{\thetable}{S\arabic{table}}

% vector variables in bold format
\def\qq{\mathbf{q}}
\def\kk{\mathbf{k}}
\def\KK{\mathbf{K}}
\def\DKK{\Delta\mathbf{K}}
\def\pp{\mathbf{p}}
\def\RR{\mathbf{R}}
\def\tt{\mathbf{t}}
\def\rr{\mathbf{r}}
\def\GG{\mathbf{G}}
\def\QQ{\mathbf{Q}}
\def\aa{\mathbf{a}}
\def\bb{\mathbf{b}}
\def\uu{\mathbf{u}}
\def\AA{\mathbf{A}}
\def\ff{\mathbf{f}}
\def\ll{\mathbf{l}}

\def\KK{\mathbf{K}}
\def\qq{\mathbf{q}}
\def\pp{\mathbf{p}}
\def\pp{\mathbf{p}}
\def\GG{\mathbf{G}}
\def\QQ{\mathbf{Q}}
\def\RR{\mathbf{R}}
\def\tt{\mathbf{t}}
\def\dd{\mathbf{d}}
\def\aa{\mathbf{a}}
\def\bb{\mathbf{b}}
\def\ee{\epsilon}
\def\CC{\mathcal{P}}
\def\UU{\mathbf{U}}

\def\BZ{{\rm BZ}}
\def\mS{{\mathcal{S}}}

\def\spin{{\varsigma}}

% second quantized operators
\def\hH{{ \hat{H} }}
\def\hrho{ \hat{\rho} }
\def\hg{\hat{g}}
\def\hS{\hat{S}}

\def\mG{{\mathcal{G}}}

\def\UC{{\hat{\Theta}}}
\def\UF{{\hat{\Sigma}}}

\def\mK{{\mathcal{K}}}
\def\pr{\prime}
% mathcal 
\def\mJ{{\mathcal{J}}}
\def\mK{{\mathcal{K}}}

\def\ie{{\it i.e.},\ }
\def\eg{{\it e.g.}\ }
\def\ea{{\it et al.}}

%%% STYLE OF PAGES NUMBERING
%\pagestyle{companion}\nouppercaseheads 
%\pagestyle{headings}
%\pagestyle{Ruled}

%\pagestyle{plain}
%\makepagestyle{plain}
%\makeevenfoot{plain}{\thepage}{}{}
%\makeoddfoot{plain}{}{}{\thepage}
%\makeevenhead{plain}{}{}{}
%\makeoddhead{plain}{}{}{}

%\maxsecnumdepth{subsection} % chapters, sections, and subsections are numbered
%\maxtocdepth{subsection} % chapters, sections, and subsections are in the Table Sof Contents

%%%---%%%---%%%---%%%---%%%---%%%---%%%---%%%---%%%---%%%---%%%---%%%---%%%

%\begin{document}

%%%---%%%---%%%---%%%---%%%---%%%---%%%---%%%---%%%---%%%---%%%---%%%---%%%
%   TITLEPAGE
%
%   due to variety of titlepage schemes it is probably better to make titlepage manually
%
%%%---%%%---%%%---%%%---%%%---%%%---%%%---%%%---%%%---%%%---%%%---%%%---%%%
\thispagestyle{empty}

\clearpage
\tableofcontents 

\clearpage
\pagenumbering{arabic}
%%%---%%%---%%%---%%%---%%%---%%%---%%%---%%%---%%%---%%%---%%%---%%%---%%%
%%%---%%%---%%%---%%%---%%%---%%%---%%%---%%%---%%%---%%%---%%%---%%%---%%%

{%%%
\sffamily
\centering
\Large

{\huge 
Supplementary Materials for "Localized interfacial phonon modes at the Electronic Axion Domain Wall"
}

\vspace{0.5cm}

%%%
}%%%

\section{Effective theory of $Na_3Bi$}

\subsection{Symmetry properties of $D_{6h}$} \label{symmd6h}

The crystal point group for $Na_3Bi$ is $D_{6h}$, generated by 6-fold rotation about the z-axis $\Hat{C}_{6z}$, 2-fold rotation about the x-axis, $\Hat{C}_{2x}$ and inversion $\Hat{I}$. $\Tilde{\Gamma}$ are the irreducible representations of the group. The $\pm$ in $\Tilde{\Gamma}^{\pm}$ denotes even (+) or odd (-) under inversion. The character table of $D_{6h}$ is given in Table \ref{d6h character} :

\begin{table}[h]
\centering
\begin{tabular}{||c c c c c c c c c c c c c||} 
 \hline
  $D_{6h}$ & $E$ & $2 C_{6z}$ & $2 C_{3z}$ & $C_{2z}$ & $2 C_{2x}$ & $2 C_{2y}$ & $I$ & $2 I C_{6z}$ & $2 I C_{3z}$ & $I C_{2z}$ & $3 I C_{2x}$ & $3 I C_{2y}$   \\ [0.5ex] 
 \hline\hline
 $\Tilde{\Gamma}_1^{\pm}$ & 1 & 1 & 1 & 1 & 1 & 1 & $\pm$ 1 & $\pm$ 1 & $\pm$ 1 & $\pm$ 1 & $\pm$ 1 & $\pm$ 1\\ 
 \hline
 $\Tilde{\Gamma}_2^{\pm}$ & 1 & 1 & 1 & 1 & -1 & -1 & $\pm$ 1 & $\pm$ 1 & $\pm$ 1 & $\pm$ 1 & $\mp$ 1 & $\mp$ 1\\
 \hline
 $\Tilde{\Gamma}_3^{\pm}$ & 1 & -1 & 1 & -1 & 1 & -1 & $\pm$ 1 & $\mp$ 1 & $\pm$ 1 & $\mp$ 1 & $\pm$ 1 & $\mp$ 1\\
 \hline
 $\Tilde{\Gamma}_4^{\pm}$ & 1 & -1 & 1 & -1 & -1 & 1 & $\pm$ 1 & $\mp$ 1 & $\pm$ 1 & $\mp$ 1 & $\mp$ 1 & $\pm$ 1\\
 \hline
 $\Tilde{\Gamma}_5^{\pm}$ & 2 & 1 & -1 & -2 & 0 & 0 & $\pm$ 2 & $\pm$ 1 & $\mp$ 1 & $\mp$ 2 & 0 & 0\\
 \hline
 $\Tilde{\Gamma}_6^{\pm}$ & 2 & -1 & -1 & 2 & 0 & 0 &$\pm$ 2 & $\mp$ 1 & $\mp$ 1 & $\pm$ 2 & 0 & 0\\
 \hline
\end{tabular}
\caption{Character Table of $D_{6h}$ group}
\label{d6h character}
\end{table}

The four-by-four Dirac matrices are defined as \begin{equation}
\begin{split}
    &\Gamma_i = \sigma_i \otimes \tau_1 , \Gamma_4 = I \otimes \tau_2 , \Gamma_5 = I \otimes \tau_3 \\ &\Gamma_{ij} = \epsilon_{ijk}\sigma_k \otimes I, \Gamma_{i4} = \sigma_i \otimes \tau_3 , \Gamma_{i5} = - \sigma_i \otimes \tau_2 \\ &\Gamma_{45} = I \otimes \tau_1; i,j = 1,2,3 
\end{split}
\end{equation}
where $I$ labels the identity matrix and $\sigma$, $\tau$ are two sets of Pauli matrices representing spin and orbital degrees of freedoms, respectively. On the basis wavefunction  $|s \uparrow\rangle , | p_{+} \uparrow \rangle , |s \downarrow\rangle , |p_{-} \downarrow \rangle$, the symmetry operators are given by 
\begin{eqnarray}
    \Hat{C}_{6z} = e^{i \frac{\pi}{6} \left( \sigma_3 \otimes (2 I - \tau_3) \right)}, \Hat{C}_{2x} = i \sigma_1 \otimes \tau_3, \Hat{I} = I \otimes \tau_3, \Hat{T} =  \left( i \sigma_2 \otimes I \right) K.
\end{eqnarray} The representations of the Dirac $\Gamma$-matrices, the momenta $\boldsymbol{k}$ and the strain tensors, as well as their behavior under time reversal $T$ are given in Table \ref{Reps}. The direct products of irreps can be found in Table \ref{direct product}. The above irreps can be even or odd ($\Tilde{\Gamma}^{\pm}$) under inversion and be $T=\pm$ under time reversal. The direct product of such irreps follow simple multiplication e.g. $\Tilde{\Gamma}^+_5 (T=+) \times \Tilde{\Gamma}^-_6 (T=-) = \Tilde{\Gamma}^-_3(T=-) + \Tilde{\Gamma}^-_4(T=-) + \Tilde{\Gamma}^-_5(T=-)$.

\begin{table}[h!]
\centering
\begin{tabular}{||c c c||} 
 \hline
  & Representation & T  \\ [0.5ex] 
 \hline\hline
 $\{\Gamma_1,\Gamma_2\}$ & $\Tilde{\Gamma}_6^-$ & \textemdash\\ 
 \hline
 $\{\Gamma_3,\Gamma_4\}$ & $\Tilde{\Gamma}_5^-$ & \textemdash\\ 
 \hline$\Gamma_5$, I $\otimes$ I & $\Tilde{\Gamma}_1^+$ & +\\ 
 \hline$\{\Gamma_{12}\},\{\Gamma_{34}\}$ & $\Tilde{\Gamma}_2^+$ & \textemdash\\ 
 \hline$\{\Gamma_{35},\Gamma_{45}\}$ & $\Tilde{\Gamma}_5^-$ & +\\ 
 \hline$\{\Gamma_{15},\Gamma_{25}\}$ & $\Tilde{\Gamma}_6^-$ & +\\ 
 \hline$\{\Gamma_{23}+\Gamma_{14},\Gamma_{31}+\Gamma_{24}\}$ & $\Tilde{\Gamma}_5^+$ & \textemdash\\ 
 \hline$\{\Gamma_{23}-\Gamma_{14},\Gamma_{31}-\Gamma_{24}\}$ & $\Tilde{\Gamma}_5^+$ & \textemdash\\ 
 \hline$\{k_x,k_y\}$ & $\Tilde{\Gamma}_5^-$ & \textemdash\\
 \hline$\{k_z\},\{k_z^3\}$ & $\Tilde{\Gamma}_2^-$ & \textemdash\\
 \hline$\{1\},\{k^2_{||}\},\{k_z^2\}$ & $\Tilde{\Gamma}_1^+$ & +\\
 \hline$\{k_x^2-k_y^2,2 k_x k_y\}$ & $\Tilde{\Gamma}_6^+$ & +\\
 \hline$\{k_x k^2_{||},k_y k^2_{||}\}$ & $\Tilde{\Gamma}_5^-$ & \textemdash\\
 \hline$\{u_{xz},u_{yz}\}$ & $\Tilde{\Gamma}_5^+$ & +\\
 \hline$\{u_{zz}\},\{u_{xx}+u_{yy}\}$ & $\Tilde{\Gamma}_1^+$ & +\\
 \hline$\{u_{xy},u_{xx}-u_{yy}\}$ & $\Tilde{\Gamma}_6^+$ & +\\
 \hline
\end{tabular}
\caption{Representations of $\Gamma$, \textbf{k} and $u_{ij}$} 
\label{Reps}
\end{table}

\begin{table}[h]
\centering
\begin{tabular}{||c c c c c c c||} 
 \hline
   & $\Tilde{\Gamma}_1$ & $\Tilde{\Gamma}_2$ & $\Tilde{\Gamma}_3$ & $\Tilde{\Gamma}_4$ & $\Tilde{\Gamma}_5$ & $\Tilde{\Gamma}_6$ \\ [0.5ex] 
 \hline\hline
 $\Tilde{\Gamma}_1$ & $\Tilde{\Gamma}_1$ & $\Tilde{\Gamma}_2$ & $\Tilde{\Gamma}_3$ & $\Tilde{\Gamma}_4$ & $\Tilde{\Gamma}_5$ & $\Tilde{\Gamma}_6$\\ 
 \hline
 $\Tilde{\Gamma}_2$ & $\Tilde{\Gamma}_2$ & $\Tilde{\Gamma}_1$ & $\Tilde{\Gamma}_4$ & $\Tilde{\Gamma}_3$ & $\Tilde{\Gamma}_5$ & $\Tilde{\Gamma}_6$\\ 
 \hline
 $\Tilde{\Gamma}_3$ & $\Tilde{\Gamma}_3$ & $\Tilde{\Gamma}_4$ & $\Tilde{\Gamma}_1$ & $\Tilde{\Gamma}_2$ & $\Tilde{\Gamma}_6$ & $\Tilde{\Gamma}_5$\\ 
 \hline
 $\Tilde{\Gamma}_4$ & $\Tilde{\Gamma}_4$ & $\Tilde{\Gamma}_3$ & $\Tilde{\Gamma}_2$ & $\Tilde{\Gamma}_1$ & $\Tilde{\Gamma}_6$ & $\Tilde{\Gamma}_5$\\ 
 \hline
 $\Tilde{\Gamma}_5$ & $\Tilde{\Gamma}_5$ & $\Tilde{\Gamma}_5$ & $\Tilde{\Gamma}_6$ & $\Tilde{\Gamma}_6$ & $\Tilde{\Gamma}_1+\Tilde{\Gamma}_2+\Tilde{\Gamma}_6$ & $\Tilde{\Gamma}_3+\Tilde{\Gamma}_4+\Tilde{\Gamma}_5$\\ 
 \hline
 $\Tilde{\Gamma}_6$ & $\Tilde{\Gamma}_6$ & $\Tilde{\Gamma}_6$ & $\Tilde{\Gamma}_5$ & $\Tilde{\Gamma}_5$ & $\Tilde{\Gamma}_3+\Tilde{\Gamma}_4+\Tilde{\Gamma}_5$ & $\Tilde{\Gamma}_1+\Tilde{\Gamma}_2+\Tilde{\Gamma}_6$\\ 
 \hline
\end{tabular}
\caption{Direct products of the irreducible representations of the $D_{6h}$ group}
\label{direct product}
\end{table}

\subsection{Effective Hamiltonian}\label{effHSuppl}

The low energy effective Hamiltonian of $Na_3Bi$ reads \cite{wang2012dirac}

\begin{equation}
    H_{Na_3 Bi} = C_0 + C_1 k_z^2 + C_2 k_{||}^2 +  \left( M_0 + M_1 k_z^2  + M_2 k_{||}^2 \right) \Gamma_5 + A_0 \left( k_x \Gamma_3 - k_y \Gamma_4 \right),
\end{equation}
in the basis $|s \uparrow\rangle , | p_{+} \uparrow \rangle , |s \downarrow\rangle , |p_{-} \downarrow \rangle$, with basis of the form $|\alpha,\sigma\rangle, \alpha=s,p_{\pm}$ represents the orbital degree of freedom and $\sigma=\uparrow,\downarrow$ is the spin degree of freedom. $C_{0,1,2}, M_{0,1,2}, A_0$ are material dependent materials \cite{wang2012dirac,dong2019electronic,de2015charting} given in Table \ref{table:material}

\begin{table} [h]
\centering
\begin{tabular}{||c c | c c | c c||}
\hline
     $C_0$ & $-0.06382 eV $ & $C_1$ &
     $8.7536 eV \mathring{A}^2$ & $C_2$ & $-8.4008 eV \mathring{A}^2 $\\ 
     \hline
 $M_0$ & -0.08686 eV &
 $M_1$ & $10.6424 eV \mathring{A}^2$ &
 $M_2$ & $10.3610 eV \mathring{A}^2$\\
 \hline
 $A_0$ & $2.4598 eV \mathring{A}$ &
 $a$ & $6.75 \times 10^6 m^2/s^2$ &
 $b$ & $3.17 \times 10^6 m^2/s^2$\\
 \hline
 $c$ & $1.01 \times 10^6 m^2/s^2$ &
 $d$ & $1.93 \times 10^6 m^2/s^2$ &
 $f$ & $11.02 \times 10^6 m^2/s^2$\\
 \hline
\end{tabular}
\caption{Materials parameters for $Na_3Bi$}
\label{table:material}
\end{table}

% $A_1$ & \\
%  $A_2$ & \\
%  $A_3$ & \\
%  $A_4$ & \\
%  $C_3$ & \\
%  $C_4$ & \\
%  $M_3$ & \\
%  $M_4$ & \\
%  $D$ & \\

%\section{Axion term}

We next introduce the Hamiltonian to describe the electron-strain coupling via symmetry construction. 
%the axion field as a strain perturbation to the above Hamiltonian. 
We define the strain tensor as $u_{ij} = \frac{1}{2}\left( \partial_i u_j + \partial_j u_i \right)$. Using the tables in Section \ref{symmd6h}, we construct the minimal Hamiltonian in the presence of a strain perturbation as \cite{winkler2003spin} 
%up to second order in momentum.  
\begin{equation}
\begin{split} 
    H_{str} &=  \left[ C_3 u_{zz} + C_4 \left( u_{xx}+u_{yy}\right)\right] \mathbb{I} + \left[ M_3 u_{zz} + M_4 \left( u_{xx}+u_{yy}\right)\right] \Gamma_5  
    + \left[ A_1 u_{zz} + A_2 \left( u_{xx}+u_{yy}\right) \right]   \left( k_x \Gamma_3 - k_y \Gamma_4 \right)\\ &+ A_3 \left(k_x  \left( u_{xx} - u_{yy}\right) + 2k_y u_{xy}   \right) \Gamma_3 - A_3 \left(  2k_x u_{xy} - k_y  \left( u_{xx} - u_{yy}\right) \right) \Gamma_4  + A_4 k_z \left( u_{xz} \Gamma_3 - u_{yz} \Gamma_4 \right) \\ &+  A_5 \left( k_x u_{yz} + k_y u_{xz} \right) \Gamma_2 + A_5 \left( k_x u_{xz} - k_y u_{yz}\right) \Gamma_1 + D k_z \left(2 u_{xy} \Gamma_2 + \left( u_{xx} - u_{yy}\right) \Gamma_1 \right) 
\end{split}
\end{equation}
where we have kept all the terms up to the linear order in both ${\bf k}$ and ${\bf u}$, as well as the order of $k_i u_{jk}$ with $i,j,k=x,y,z$. The Dirac points of $H_{Na_3 Bi}$ are located at the momenta $ \left(0,0,a k_c \right)$, where $k_c = \sqrt{-\frac{M_0}{M_1}}$ and $a=\pm$ are dubbed two "valleys" below. We project both the Hamiltonians $H_{Na_3 Bi} + H_{str}$ into the subspace formed by four eigen-states  at the Dirac points of $H_{Na_3 Bi}$ at 
$\left(0,0,a k_c\right)$. We keep all the terms up to the linear order in ${\bf k}$. 
The total effective Hamiltonian after the projection is given by $\Tilde{H} = \sum_a \Tilde{H}_a$. The valley dependent term is given by 
\begin{eqnarray}
    && \Tilde{H}_a = \left[ C_3 u_{zz} + C_4 \left( u_{xx}+u_{yy}\right)\right] \mathbb{I} + A_0 \left[  k_x + a \frac{A_4}{A_0} k_c u_{xz} \right]  \Gamma_3 - A_0 \left[  k_y + a \frac{A_4}{A_0} k_c u_{yz} \right] \Gamma_4 \nonumber \\ 
    &&+ 2 a  M_1 k_c \left[ k_z + \frac{a}{2 M_1 k_c} \left( M_3 u_{zz} + M_4 \left( u_{xx}+u_{yy}\right) \right) \right] \Gamma_5 + 2 a D k_c  u_{xy} \Gamma_2 \nonumber \\  &&+ D k_c a  \left(u_{xx} - u_{yy} \right) \Gamma_1 . 
\end{eqnarray}
In $\Tilde{H_a}$, the $z$ component of Fermi velocity given by $2 a M_1 k_c$ is dependent on the valley index $a$. In order to remove this valley dependence, we perform a unitary transformation for $H_{a=+}$ with $U=\Gamma_{15}$, but not for $H_{a=-}$, since $H_{a=-}$ and $H_{a=+}$ are independent of each other. 
%, this unitary transformation does not effect $H_{a=-}$ i.e. 
Thus, $H_{a=-} = \Tilde{H}_{a=-}$ and $H_{a=+} = U \Tilde{H}_{a=+} U^{-1}$. We can rewrite the valley dependent Hamiltonian as

\begin{equation}\label{eq:H_valley_Na3Bi}
\begin{split}
    H_a &= \left[ C_3 u_{zz} + C_4 \left( u_{xx}+u_{yy}\right)\right] \mathbb{I} + A_0 \left[  k_x + a \frac{A_4}{A_0} k_c u_{xz} \right]  \Gamma_3 - A_0 \left[  k_y + a \frac{A_4}{A_0} k_c u_{yz} \right] \Gamma_4 \\ & - 2 M_1 k_c \left[ k_z + \frac{a}{2 M_1 k_c} \left( M_3 u_{zz} + M_4 \left( u_{xx}+u_{yy}\right) \right) \right] \Gamma_5 + D k_c \Big[ \left(u_{xx} - u_{yy} \right) \Gamma_1 - 2 a u_{xy}  \Gamma_2  \Big]  
\end{split}
\end{equation}

The strain perturbation has induced a pseudo-gauge field of the form $(A_0^{pse}, \Vec{A}^{pse})$ with  

\begin{equation}
\begin{split} \label{Psuedo gauge field}
    A_0^{pse} &= C_3 u_{zz} + C_4 \left(u_{xx} + u_{yy} \right)  \\ \Vec{A}^{pse} &= -a \Big[ - A_4 k_c u_{xz} , A_4 k_c u_{yz} , M_3 u_{zz} + M_4 \left(u_{xx} + u_{yy} \right)    \Big]
\end{split}
\end{equation}

\subsection{Effective action $S_{ax}$} \label{effaction}

The above Dirac type of Hamiltonian may be re-written as a Lagrangian density 

\begin{equation} \label{eq:L_valley_Na3Bi}
    \mathcal{L} = \sum_{a=\pm} \mathcal{L}_a = \sum_{a=\pm} \Bar{\psi}_a \left[ i \slashed{\partial} - e \slashed{A}_a^{pse} + |m| \left( \cos\Phi_a I + \sin\Phi_a \gamma^5 \right) \right] \psi_a
\end{equation}
where $\Bar{\psi}_a = \psi^\dagger \gamma^0$ , $\slashed{\partial}_a = \gamma^\mu \partial_{\mu,a}$, $\slashed{A}_a^{pse} = \gamma^\mu A_{\mu,a}^{pse}$, $\partial_{0,a} = \partial_t$, $\partial_{1,a} = A_0 \partial_x$, $\partial_{2,a} = - A_0 \partial_y$, $\partial_{3,a} =  - 2 M_1 k_c \partial_z$. The mass term is given by $|m| = D k_c \left((2u_{xy})^2 + (u_{xx}-u_{yy})^2\right)^{1/2}$ . We have $\gamma^0 = \Gamma_1, \gamma^1 = \Gamma_1 \Gamma_3 = i \Gamma_{13}, \gamma^2 = \Gamma_1 \Gamma_4 = i \Gamma_{14}, \gamma^3 = \Gamma_1 \Gamma_5 = i \Gamma_{15}, \gamma^5 = i \gamma^0 \gamma^1 \gamma^2 \gamma^3 = -i \sigma_3 \otimes I$, $\Gamma_2=-i\gamma^0\gamma^5 = \sigma_2 \otimes \tau_1$, 
\begin{align}\label{Phi_valley}
    \Phi_a = a \theta; \: \: \: \: \cot\theta = \frac{u_{xx}-u_{yy}}{2 u_{xy}}.
\end{align}

The corresponding Hamiltonian form of the above Lagrangian is 
\begin{equation}\label{HDiracGen}
    H = \sum_{a=1}^2 H_a = \sum_{a=1}^2 \psi^\dagger_a \Big[A_0^{pse} + \gamma^0 \gamma^j \left( -i \partial_j +  A_j^{pse}\right) - |m| \gamma^0 \left( \cos\Phi_a I + \sin\Phi_a \gamma^5 \right) \Big] \psi_a. 
\end{equation}
The above Hamiltonian can be compared with the Hamiltonian $H_a(\{u\}=0)$ in Eq. (\ref{eq:H_valley_Na3Bi}) to determine the forms of $\gamma$'s in terms of $\Gamma$'s. Three linear-momentum terms can be uniquely fixed by requiring 
\begin{align}
    \gamma^0 \gamma^1 &= \Gamma_3 = \sigma_3 \otimes \tau_1 \\
    \gamma^0 \gamma^2 &= \Gamma_4 = I \otimes \tau_2 \\
    \gamma^0 \gamma^3 &= \Gamma_5 = I \otimes \tau_3. 
\end{align}    

There are still two possible choices to assign the two mass terms when comparing Eq(\ref{HDiracGen}) with Eq. (\ref{eq:H_valley_Na3Bi}). As the mirror symmetry about $yz$ plane, $m_x$, is preserved for the model Hamiltonian of $Na_3 Bi$, two mass terms in Eqs.(\ref{HDiracGen}) and (\ref{eq:H_valley_Na3Bi}) should be transformed in the same way. 
In the $Na_3Bi$ basis of the Hamiltonian (\ref{eq:H_valley_Na3Bi}), we have $\Hat{m}_x = \Hat{I} C_{2x} = (I \otimes \tau_3)(i\sigma_1 \otimes \tau_3) = i \sigma_1 \otimes I$. Under $\Hat{m}_x$, $\Gamma_1$ transforms as:

\begin{equation}
    m_x^{-1} \Gamma_1 m_x = -i (\sigma_1 \otimes I) (\sigma_1 \otimes \tau_1) i (\sigma_1 \otimes I) = \Gamma_1
\end{equation}

Similarly, $\Gamma_2$ transforms as:

\begin{equation}
    m_x^{-1} \Gamma_2 m_x = -i (\sigma_1 \otimes I) (\sigma_2 \otimes \tau_1) i (\sigma_1 \otimes I) = - \Gamma_2
\end{equation}
In the spinor basis of the Dirac equation, we have $\Hat{I} \equiv \gamma^0$ and $\Hat{C}_{2x} = e^{-i \frac{\pi}{2} \sigma_{23}} = - i\sigma_{23} \equiv S $ where $\sigma_{23} = \frac{i}{2} [\gamma^2, \gamma^3]$. We have,
\begin{align}
    \hat{m}_x^{-1} \gamma^0 \hat{m}_x & = \left( \Hat{I} \Hat{C}_{2x} \right)^{-1} \gamma^0 \left( \Hat{I} \Hat{C}_{2x} \right) = \left( \gamma^0 S \right)^{-1} \gamma^0 \left( \gamma^0 S \right) = S^{-1} \gamma^0 \gamma^0 \gamma^0 S = S^{-1} \gamma^0  S \nonumber \\
    &= \left( \cos \frac{\pi}{2} + i \sin \frac{\pi}{2} \sigma_{23} \right) \gamma^0 \left( \cos \frac{\pi}{2} - i \sin \frac{\pi}{2} \sigma_{23} \right) = \sigma_{23} \gamma^0 \sigma_{23} \nonumber \\
    &= \frac{i}{2} [\gamma^2, \gamma^3] \gamma^0 \frac{i}{2} [\gamma^2, \gamma^3] \nonumber = -\frac{1}{4} \left( \gamma^2 \gamma^3 - \gamma^3 \gamma^2 \right) \gamma^0 \left( \gamma^2 \gamma^3 - \gamma^3 \gamma^2 \right) \nonumber \\
    &= -\frac{1}{4} \gamma^0 \left( \gamma^2 \gamma^3 - \gamma^3 \gamma^2 \right)^2 = -  \frac{1}{4} \gamma^0 \left( -4 I\right)  \nonumber \\ &= \gamma^0 
\end{align}
Since $\gamma_0$ is invariant under $\Hat{I} \Hat{C}_{2x}$, we must have $\gamma^0 = \Gamma_1$. Similarly, one can show $\hat{m}_x^{-1} \gamma^0 \gamma^5 \hat{m}_x = - \gamma^0 \gamma^5$ so that $\gamma^0 \gamma^5=\Gamma_2$.

With the above Lagrangian Eq. (\ref{eq:L_valley_Na3Bi}), we can derive the total effective action as $S_{ax} = \sum_a S_{eff,a}$ by integrating out the Dirac fermions  \cite{yu2021dynamical} via the path integral form
\begin{equation}
    e^{i S_{eff,a}}  = \int D\Bar{\psi}_a D\psi_a exp\left[ i \int dt d^3r \mathcal{L}_a \right],
\end{equation}
where
\begin{equation}
    S_{eff,a} = \frac{1}{32 \pi^2} \int dt d^3 r \Phi_a \epsilon^{\mu \nu \rho \delta} F_{\mu \nu,a}^{pse} F_{\rho \delta,a}^{pse}
\end{equation}
and $F_{\mu \nu, a}^{pse} = \partial_{\mu} A_{\nu,a}^{pse} - \partial_{\nu} A_{\mu,a}^{pse}$. Since we only discuss the pseudo-gauge field, we drop the upper index "pse". We define $\Tilde{C} = C_3 u_{zz} + C_4 \left( u_{xx} + u_{yy} \right) $, $\Tilde{M} = M_3 u_{zz} + M_4 \left( u_{xx} + u_{yy} \right) $, $\partial_1 = A_0 \partial_x$, $\partial_2 = -A_0 \partial_y$ and $\partial_3 = -2 M_1 k_c \partial_z$, and find
\begin{align}
    F_{01,a} &= \Dot{A}_{1,a} - \partial_1 A_{0,a} = \Big[a A_4 k_c \Dot{u}_{xz} - A_0 \partial_x \Tilde{C} \Big]\nonumber \\
    F_{02,a} &= \Dot{A}_{2,a} - \partial_2 A_{0,a} = \Big[-a A_4 k_c \Dot{u}_{yz} + A_0 \partial_y \Tilde{C} \Big]\nonumber \\
    F_{03,a} &= \Dot{A}_{3,a} - \partial_3 A_{0,a} = \Big[-a \Dot{M} + 2 M_1 k_c \partial_z \Tilde{C} \Big]\nonumber \\
    F_{12,a} &= \partial_1 A_{2,a} - \partial_2 A_{1,a} = -a  A_0 A_4 k_c \Big[ \partial_x u_{yz} - \partial_y u_{xz} \Big] \nonumber \\
    F_{23,a} &= \partial_2 A_{3,a} - \partial_3 A_{2,a} = a \Big[ A_0 \partial_y \Tilde{M} - 2 A_4 M_1 k_c^2 \partial_z u_{yz}\Big] \nonumber \\
    F_{13,a} &= \partial_1 A_{3,a} - \partial_3 A_{1,a} = -a \Big[A_0 \partial_x \Tilde{M} - 2 A_4 M_1 k_c^2 \partial_z u_{xz} \Big] . 
\end{align}
With Eq. (\ref{Phi_valley}) and Eq(\ref{Psuedo gauge field}) for $\Phi_a$ and $A_{\mu,a}^{pse}$, we can obtain the total effective action 
\begin{equation}\label{S1}
    S_{ax} =  \int dt d^3r \left( \frac{1}{2 A_0^2 M_1 k_c} \right) \frac{\theta}{2 \pi^2} \Bigg[ \partial_x \Tilde{C} \left( J \partial_z u_{yz} - A_0^2 \partial_y \Tilde{M} \right) + \partial_y \Tilde{C} \left( -J \partial_z u_{xz} + A_0^2 \partial_x \Tilde{M} \right) + \partial_z \Tilde{C} \left( - J \partial_x u_{yz} + J \partial_y u_{xz} \right) \Bigg]
\end{equation}
where $J = - 2 A_0 A_4 M_0 $. One notices that all the terms that depend on time derivative contain the valley index $a$ because of time reversal symmetry, so they vanish when summing over $a$. We can write $S_{ax}$ as a sum of total derivatives as
\begin{equation}
    S_{ax} =  \int dt d^3r \left( \frac{1}{2 A_0^2 M_1 k_c} \right) \frac{\theta}{2 \pi^2} \Bigg[ \partial_x \Big[\Tilde{C}  \left(J \partial_z u_{yz} - A_0^2\partial_y \Tilde{M} \right)\Big] + \partial_y \Big[\Tilde{C} \left( -J \partial_z u_{xz} + A_0^2 \partial_x \Tilde{M} \right)\Big] +  \partial_z \Big[\Tilde{C} \left( - J \partial_x u_{yz}  + J \partial_y u_{xz} \right)\Big] \Bigg]
\end{equation}
We assume periodic conditions along the $x$ and $z$ directions, as such being total derivatives, these contributions vanish. We are left with the $y$ dependence of $\theta$ and redefine $\frac{\theta}{2 \pi^2} \left( \frac{1}{2 A_0^2 M_1 k_c} \right) \rightarrow \theta $ so that the effective action is re-written as
\begin{equation} \label{Sxz1}
    S_{ax} = -  \int dt dx dy dz \frac{\partial \theta}{\partial y} \Big[\Tilde{C} \left( -J \partial_z u_{xz} + A_0^2 \partial_x \Tilde{M} \right)\Big]
\end{equation}

\subsection{Symmetries of $S_{ax}$} \label{actionsymm}
The parameter $\theta$ in the effective action is originated from the valley axion field $\Phi_a$, and therefore the symmetry properties of $\theta$ need not necessarily follow the symmetry properties of the standard axion term. In this section, we summarize the symmetries of the effective action $S_{ax}$ as follows. Under six-fold rotation, $C_{6z}$, the effective action $S_{ax}$ is invariant. Since the strain tensors as well as the partial derivatives are even under time-reversal, $S_{ax}$ is invariant under time-reversal symmetry. Under inversion, the strain tensors are even whereas the partial derivatives are odd. But as can be seen in Eq(\ref{S1}), the partial derivatives occur in pairs. So $S_{ax}$ is invariant under inversion as well. $S_{ax}$ is also invariant under two-fold rotation $C_{2x}$ and $x$-reflection $m_x$.

\subsubsection{Six-fold rotation: $\Hat{C}_{6z}$}
Under $C_{6z}$, which is defined by the following transformation (for $\phi =  2\pi/6$)
\begin{equation}
    \begin{pmatrix}
        x' \\
        y' \\
        z' \\
    \end{pmatrix}
    = \begin{pmatrix}
        \cos\phi & -\sin \phi &0 \\
        \sin \phi & \cos\phi & 0 \\
        0 & 0 & 1 \\
    \end{pmatrix} 
    \begin{pmatrix}
        x \\
        y \\
        z \\
    \end{pmatrix},
\end{equation}
we have
\begin{align}
    \partial_x' &= \cos \phi \partial_x - \sin \phi \partial_y ; \partial_y' = \sin \phi \partial_x + \cos \phi \partial_y; \partial_z' = \partial_z \nonumber \\  
    u_{xz}' &= \cos \phi u_{xz} - \sin \phi u_{yz} ; u_{yz}' = \sin \phi u_{xz} + \cos \phi u_{yz}; u_{zz}' = u_{zz} \nonumber \\
    u_{xx}' - u_{yy}' &= \cos 2\phi \left( u_{xx} - u_{yy} \right) - \sin 2\phi \left(2 u_{xy} \right); 2 u_{xy}' = \sin 2 \phi  \left( u_{xx} - u_{yy} \right) + \cos 2\phi \left(2 u_{xy} \right) \nonumber  \\
    u_{xx}' + u_{yy}' &= u_{xx} + u_{yy}; \Tilde{C}' = \Tilde{C} ; \Tilde{M}' = \Tilde{M}. \nonumber
\end{align}
which lead to
\begin{equation}
    \cot \theta' = \frac{u_{xx}'-u_{yy}'}{2 u_{xy}'} = \frac{\left( u_{xx} - u_{yy}\right) \cot 2\phi   - \left( 2 u_{xy}\right) }{ \left( u_{xx} - u_{yy}\right) + 2 u_{xy} \cot 2\phi } = \cot \left( \theta + 2\pi/3\right)
\end{equation}
and
\begin{eqnarray}
        &&\partial_x' \Tilde{C}' \left( J \partial_z' u_{yz}' - \partial_y' \Tilde{M}' \right) + \partial_y' \Tilde{C}' \left( -J \partial_z' u_{xz}' + \partial_x' \Tilde{M}' \right) + \partial_z' \Tilde{C}' \left( - J \partial_x' u_{yz}' + J \partial_y' u_{xz}' \right) \nonumber \\
        &&= \left( \cos \phi \partial_x \Tilde{C} - \sin \phi \partial \Tilde{C} \right) \left( J \sin\phi \partial_z u_{xz} + J \cos\phi \partial_z u_{yz} - \sin\phi \partial_x \Tilde{M} - \cos\phi \partial_y \Tilde{M} \right) \nonumber \\
        && - \left(\sin\phi \partial_x \Tilde{C} + \cos\phi \partial_y \Tilde{C} \right) \left(J \cos\phi \partial_z u_{xz} - J \sin\phi \partial_z u_{yz} - \cos\phi \partial_x \Tilde{M} +\sin\phi \partial_y \Tilde{M} \right) \nonumber \\
        && - \partial_z \Tilde{C} \left( J \left( \cos\phi \partial_x - \sin\phi \partial_y\right) \left( \sin\phi u_{xz}+ \cos\phi u_{yz}\right) - J \left(\sin\phi \partial_x + \cos\phi \partial_y \right)\left(\cos\phi u_{xz} - \sin\phi u_{yz} \right)   \right) \nonumber \\
         &&= \partial_x \Tilde{C} \left( J \partial_z u_{yz} - \partial_y \Tilde{M} \right) + \partial_y \Tilde{C} \left( -J \partial_z u_{xz} + \partial_x \Tilde{M} \right) + \partial_z \Tilde{C} \left( - J \partial_x u_{yz} + J \partial_y u_{xz} \right) . 
\end{eqnarray}

Therefore, under $C_{6z}$, $ \theta\rightarrow \theta + 2\pi/3$ i.e. 
\begin{align}
    S_{ax}' &=  \int dt d^3r \left( \frac{1}{2 A_0^2 M_1 k_c} \right) \frac{\theta + 2 \pi/3}{2 \pi^2} \Bigg[ \partial_x \Tilde{C} \left( J \partial_z u_{yz} - A_0^2 \partial_y \Tilde{M} \right) + \partial_y \Tilde{C} \left( -J \partial_z u_{xz} + A_0^2 \partial_x \Tilde{M} \right) + \partial_z \Tilde{C} \left( - J \partial_x u_{yz} + J \partial_y u_{xz} \right) \Bigg] \nonumber \\
    &= \int dt d^3r \left( \frac{1}{2 A_0^2 M_1 k_c} \right) \frac{\theta }{2 \pi^2} \Bigg[ \partial_x \Tilde{C} \left( J \partial_z u_{yz} - A_0^2 \partial_y \Tilde{M} \right) + \partial_y \Tilde{C} \left( -J \partial_z u_{xz} + A_0^2 \partial_x \Tilde{M} \right) + \partial_z \Tilde{C} \left( - J \partial_x u_{yz} + J \partial_y u_{xz} \right) \Bigg] \nonumber \\
    &+ \int dt d^3r \left( \frac{1}{2 A_0^2 M_1 k_c} \right) \frac{2 \pi /3 }{2 \pi^2} \Bigg[ \partial_x \Tilde{C} \left( J \partial_z u_{yz} - A_0^2 \partial_y \Tilde{M} \right) + \partial_y \Tilde{C} \left( -J \partial_z u_{xz} + A_0^2 \partial_x \Tilde{M} \right) + \partial_z \Tilde{C} \left( - J \partial_x u_{yz} + J \partial_y u_{xz} \right) \Bigg]\nonumber \\
    & = S_{ax}
\end{align}
In the second equality above, the second term is a total derivative and vanishes, thereby leaving $S_{ax}$ invariant. 

\subsubsection{Two-fold rotation $\Hat{C}_{2x}$}
Under $C_{2x}$ that transforms as
\begin{equation}
    \begin{pmatrix}
        x' \\
        y' \\
        z' \\
    \end{pmatrix}
    = \begin{pmatrix}
        1 & 0 &0 \\
        0 & -1 & 0 \\
        0 & 0 & -1 \\
    \end{pmatrix} 
    \begin{pmatrix}
        x \\
        y \\
        z \\
    \end{pmatrix}
\end{equation}
we have
\begin{align}
    \partial_x' &= \partial_x ; \partial_{y,z}' = - \partial_{y,z} \nonumber \\
    u_{xz}' &= -u_{xz} ; u_{yz}' = u_{yz}; u_{xx,yy,zz}' = u_{xx,yy,zz} \nonumber \\
    2u_{xy}' &= - 2u_{xy} ; u_{xx}' \pm u_{yy}' = u_{xx} \pm u_{yy} \nonumber \\
    \Tilde{C}' &= \Tilde{C} ; \Tilde{M}' = \Tilde{M},
\end{align}
and \begin{equation}
    \cot \theta' = \frac{u_{xx}'-u_{yy}'}{2 u_{xy}'} = - \frac{u_{xx}-u_{yy}}{2 u_{xy}} =  \cot \left( -\theta \right).
\end{equation}
 Since
\begin{eqnarray}
        &&\partial_x' \Tilde{C}' \left( J \partial_z' u_{yz}' - \partial_y' \Tilde{M}' \right) + \partial_y' \Tilde{C}' \left( -J \partial_z' u_{xz}' + \partial_x' \Tilde{M}' \right) + \partial_z' \Tilde{C}' \left( - J \partial_x' u_{yz}' + J \partial_y' u_{xz}' \right) \nonumber \\
         &&= - \Big[ \partial_x \Tilde{C} \left( J \partial_z u_{yz} - \partial_y \Tilde{M} \right) + \partial_y \Tilde{C} \left( -J \partial_z u_{xz} + \partial_x \Tilde{M} \right) + \partial_z \Tilde{C} \left( - J \partial_x u_{yz} + J \partial_y u_{xz} \right) \Big],
\end{eqnarray}
 under $C_{2x}$, $ S_{ax} \rightarrow S_{ax}$. 

\subsubsection{Reflection $m_x$}
Under $m_x$ that transforms as
\begin{equation}
    \begin{pmatrix}
        x' \\
        y' \\
        z' \\
    \end{pmatrix}
    = \begin{pmatrix}
        -1 & 0 &0 \\
        0 & 1 & 0 \\
        0 & 0 & 1 \\
    \end{pmatrix} 
    \begin{pmatrix}
        x \\
        y \\
        z \\
    \end{pmatrix},
\end{equation}
we have
\begin{align}
    \partial_x' &= - \partial_x ; \partial_{y,z}' = \partial_{y,z} \nonumber \\
    u_{xz}' &= -u_{xz} ; u_{yz}' = u_{yz}; u_{xx,yy,zz}' = u_{xx,yy,zz} \nonumber \\
    2u_{xy}' &= - 2u_{xy} ; u_{xx}' \pm u_{yy}' = u_{xx} \pm u_{yy} \nonumber \\
    \Tilde{C}' &= \Tilde{C} ; \Tilde{M}' = \Tilde{M}
\end{align}
and \begin{equation}
    \cot \theta' = \frac{u_{xx}'-u_{yy}'}{2 u_{xy}'} = - \frac{u_{xx}-u_{yy}}{2 u_{xy}} =  \cot \left( -\theta \right).
\end{equation}
 Since
\begin{eqnarray}
        &&\partial_x' \Tilde{C}' \left( J \partial_z' u_{yz}' - \partial_y' \Tilde{M}' \right) + \partial_y' \Tilde{C}' \left( -J \partial_z' u_{xz}' + \partial_x' \Tilde{M}' \right) + \partial_z' \Tilde{C}' \left( - J \partial_x' u_{yz}' + J \partial_y' u_{xz}' \right) \nonumber \\
         &&= - \Big[ \partial_x \Tilde{C} \left( J \partial_z u_{yz} - \partial_y \Tilde{M} \right) + \partial_y \Tilde{C} \left( -J \partial_z u_{xz} + \partial_x \Tilde{M} \right) + \partial_z \Tilde{C} \left( - J \partial_x u_{yz} + J \partial_y u_{xz} \right) \Big],
\end{eqnarray}
 under $m_x$, $ S_{ax} \rightarrow S_{ax}$. 

\subsection{Bulk action} \label{sectiond6h}
We first consider the bulk elastic wave in a hexagonal crystal lattice with $D_{6h}$ symmetry group.
The most general form of the elastic wave action is  
\begin{equation}
    S_0 = \int dt d^3r \Big[ \frac{1}{2} \rho_{ij} \partial_t u_i \partial_t u_j - F_0  \Big] ; \ \ F_0 = \frac{1}{2} \lambda_{ijkl} u_{ij} u_{kl} \label{S0general}
\end{equation}
We assume the bulk to be uniform, i.e. $\rho_{ij} = \rho \delta_{ij}$ and denote the free energy as $ F_0 = \frac{1}{2} \lambda_{ijkl} u_{ij} u_{kl}$. We follow Ref. \cite{landau1986theory} to construct the free energy for a hexagonal crystal by symmetry. We make a coordinate transformation, i.e. $\xi = x+iy, \eta = x-iy$. As a result, the $n$-fold rotation along the $z$-axis, $\Hat{C}_{nz}$, gives $\Hat{C}_{nz} \xi = \xi e^{i2\pi/n}, \Hat{C}_{nz} \eta = \xi e^{-i2\pi/n}$. The derivatives transform as $\partial_\xi = \frac{1}{2} \left(\partial_x - i\partial_y \right), \partial_\eta = \frac{1}{2} \left(\partial_x + i \partial_y \right)$. The strain tensors transform as
\begin{equation} \label{coord}
    \begin{split}
        u_{\xi\xi} &= u_{xx} - u_{yy} + 2 i u_{xy} , u_{\eta\eta} = u_{xx} - u_{yy} - 2 i u_{xy} \\
        u_{\xi z} &= u_{xz} + i u_{yz}, u_{\eta z} = u_{xz} - i u_{yz} , u_{\xi \eta} = u_{xx} + u_{yy} 
    \end{split}. 
\end{equation}
We need $F_0$ to be constrained by the generators of the $D_{6h}$ group. It turns out that the only constraint comes from the 6-fold rotation operator $\Hat{C}_{6z}$. In order for $F_0$ to be invariant under $\Hat{C}_{6z}$, the indices $\xi$ and $\eta$ must occur in pairs. For instance, a term of the form $\sim u_{\xi \xi} u_{\xi \eta}$ will break $C_{6z}$, which can be derived using Eq(\ref{coord}). Below, we write down all the possible terms and their combinatorial factors e.g. the term $\frac{1}{2} \lambda_{\xi\eta z z} u_{\xi \eta} u_{zz}$ would have a factor of 4 because $\lambda_{\xi \eta zz} = \lambda_{\eta \xi zz} = \lambda_{zz \eta \xi } = \lambda_{zz \xi \eta } $. The form of $F_0$ is  
\begin{equation}
    \begin{split}
        F_0 &=\frac{1}{2} \lambda_{zzzz} u_{zz}^2 + 4 \frac{1}{2} \lambda_{\xi\eta\xi\eta} u_{\xi\eta}^2 + 2 \frac{1}{2} \lambda_{\xi\xi\eta\eta} u_{\xi\xi} u_{\eta \eta} +4 \frac{1}{2} \lambda_{\xi\eta z z} u_{\xi \eta} u_{zz} + 8 \frac{1}{2} \lambda_{\xi z\eta z} u_{\xi z} u_{\eta z} \\
        &=\frac{1}{2} \lambda_{zzzz} u_{zz}^2 + 2 \lambda_{\xi\eta\xi\eta} u_{\xi\eta}^2 + \lambda_{\xi\xi\eta\eta} u_{\xi\xi} u_{\eta \eta} + 2 \lambda_{\xi\eta z z} u_{\xi \eta} u_{zz} + 4 \lambda_{\xi z\eta z} u_{\xi z} u_{\eta z}
    \end{split}
\end{equation}

Using Eq(\ref{coord}), we transform the above free energy back into Cartesian coordinates and find
\begin{equation}
    \begin{split}
        F_0 &= \frac{1}{2} \lambda_{zzzz} u_{zz}^2 + \frac{1}{2} \lambda_{xxxx} u_{xx}^2 + \frac{1}{2} \lambda_{yyyy} u_{yy}^2 + \lambda_{xxyy} u_{xx} u_{yy} + \lambda_{xxzz} u_{xx} u_{zz} \\&+ \lambda_{yyzz} u_{yy} u_{zz} + 2 \lambda_{xyxy} u_{xy}^2 + 2 \lambda_{xzxz} u_{xz}^2 + 2 \lambda_{yzyz} u_{yz}^2
    \end{split}. 
\end{equation}

We define
\begin{equation} \label{5const}
    \begin{split}
        \lambda_{xxxx} &= \lambda_{yyyy} = a+b , \lambda_{xxyy} = a-b , \lambda_{xyxy} = b , \lambda_{xxzz} = \lambda_{yyzz} = c \\ \lambda_{xzxz} &= \lambda_{yzyz} = d , \lambda_{zzzz} = f
    \end{split}
\end{equation}

which correspond to \begin{equation}
    \begin{split}
        4 \lambda_{\xi \eta \xi \eta} = a , 2 \lambda_{\xi \xi \eta \eta} = b , 2 \lambda_{\xi \eta z z} = c, 2 \lambda_{\xi z \eta z} = d, \lambda_{zzzz} = f
    \end{split}
\end{equation}

For $Na_3 Bi$, the elastic tensor \cite{de2015charting,dong2019electronic} can be written in a matrix form
\begin{equation}
    \Lambda_{theoretical}  = \begin{pmatrix} 
     \lambda_{xxxx} & \lambda_{xxyy} & \lambda_{xxzz} & 0 & 0 & 0 \\
     \lambda_{yyxx} & \lambda_{yyyy} & \lambda_{yyzz} & 0 & 0 & 0 \\
     \lambda_{zzxx} & \lambda_{zzyy} & \lambda_{zzzz} & 0 & 0 & 0 \\
     0 & 0 & 0 & \lambda_{yzyz} & 0 & 0 \\
     0 & 0 & 0 & 0 & \lambda_{xzxz} & 0 \\
     0 & 0 & 0 & 0 & 0 & \lambda_{xyxy} \\
    \end{pmatrix} 
   = \begin{pmatrix} 
     a+b & a-b & c & 0 & 0 & 0 \\
     a-b & a+b & c & 0 & 0 & 0 \\
     c & c & f & 0 & 0 & 0 \\
     0 & 0 & 0 & d & 0 & 0 \\
     0 & 0 & 0 & 0 & d & 0 \\
     0 & 0 & 0 & 0 & 0 & b \\
    \end{pmatrix},
\end{equation}

The values of the paramters $a,b,c,d,f$ can be found in Table \ref{table:material} according to Refs. \cite{de2015charting,dong2019electronic}. 

\begin{figure*}
 \includegraphics[width=\textwidth]{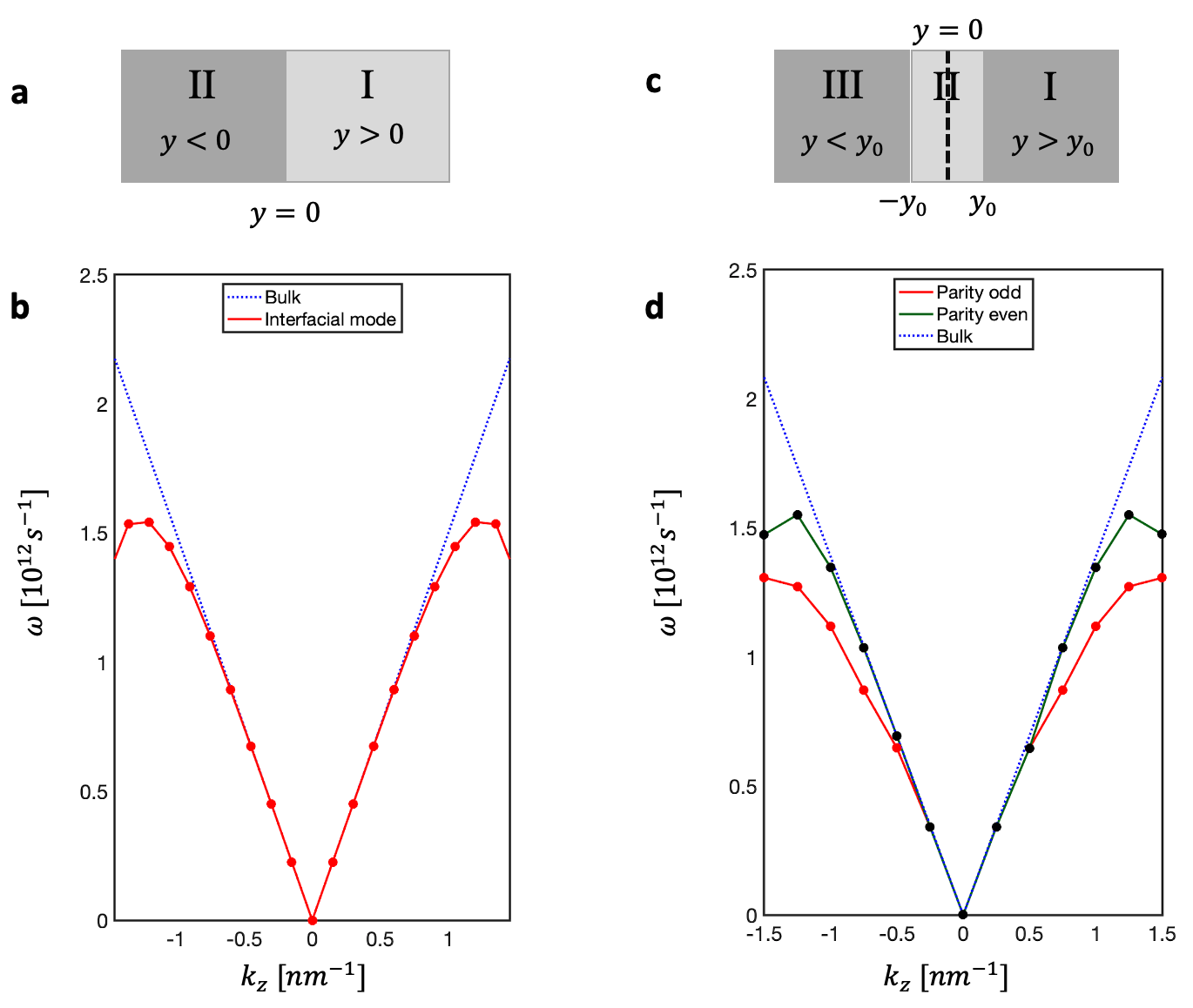}
 \caption{(a) Regions I and II for \textit{Case I} (b) Interfacial mode (red) frequency as a function of $k_z$ (for $k_x=0$) and $c_0 = 1.5 \times 10^{3} m/s$, $r=0.8$ in units of $10^{-12} m^4/s^2$ (c) Regions I, II and III for \textit{Case II}. (d) Frequency of interfacial modes (even and odd parities) for $k_x=0$ and $\frac{\theta_0 s}{y_0} = 5$ in units of $10^{-9} m^4/s^2$  with $a=6.75, b=3.17$ in units of $10^6 m^2/s^2$}
 \label{fig:S1}
 \end{figure*}

\subsection{Equation of motion with the Axion term}\label{EOMaxion}

In this section, we derive the full equation of motion including the axion contribution. Here we rewrite the axion action $S_{ax}$ into a general form $S_{ax} = S_{11} +S_{12} + S_{13}$
where (assuming Einstein's summation convention)
\begin{align}
    S_{11} &= -\int d^3 r \frac{\partial \theta}{\partial y} \Big[ \xi_{ijkls} \partial_k u_i \partial_l \partial_s u_j \Big] \label{S11general}\\ 
    S_{12} &= -\int d^3 r \frac{\partial \theta}{\partial y} \Big[ \eta_{ijkl} \partial_k u_i \partial_y \partial_l u_j \Big] \label{S12general}\\ 
    S_{13} &= -\int d^3 r \frac{\partial \theta}{\partial y} \Big[ \phi_{ijk} \partial_y u_i \partial_y \partial_k u_j \Big] \label{S13general}.
\end{align}

By expanding Eq(\ref{Sxz1}), we find
\begin{align} \label{S1xz}
    S_1 &= \int d^3 r \frac{\partial \theta}{\partial y} \Big[ -\frac{J}{2} C_3 \partial_z u_z \partial_z^2 u_x - \frac{J}{2} C_3 \partial_z u_z \partial_x \partial_z u_z - \frac{J}{2} C_4 \partial_x u_x \partial_z^2 u_x - \frac{J}{2} C_4 \partial_x u_x \partial_x \partial_z u_z \nonumber \\ &- \frac{J}{2} C_4 \partial_y u_y \partial_z^2 u_x - \frac{J}{2} C_4 \partial_y u_y \partial_x \partial_z u_z +A_0^2 M_3 C_3 \partial_z u_z \partial_x \partial_z u_z + A_0^2 C_4 M_3 \partial_x u_x \partial_x \partial_z u_z \nonumber \\
    &+ A_0^2 C_4 M_3 \partial_y u_y \partial_x \partial_z u_z + A_0^2 C_3 M_4 \partial_z u_z \partial_x^2 u_x + A_0^2 C_4 M_4 \partial_x u_x \partial_x^2 u_x + A_0^2 C_4 M_4 \partial_y u_y \partial_x^2 u_x \nonumber \\
    &+ A_0^2 C_3 M_4 \partial_z u_z \partial_x \partial_y u_y 
    + A_0^2 C_4 M_4 \partial_x u_x \partial_x \partial_y u_y + A_0^2 C_4 M_4 \partial_y u_y \partial_x \partial_y u_y \Big] \nonumber \\
    &= \int d^3 r \frac{\partial \theta}{\partial y} \Big[ -\frac{J}{2} C_3 \partial_z u_z \partial_z^2 u_x  - \frac{J}{2} C_4 \partial_x u_x \partial_x \partial_z u_z - \frac{J}{2} C_4 \partial_y u_y \partial_z^2 u_x - \frac{J}{2} C_4 \partial_y u_y \partial_x \partial_z u_z \nonumber \\ &+ A_0^2 C_4 M_3 \partial_x u_x \partial_x \partial_z u_z + A_0^2 C_4 M_3 \partial_y u_y \partial_x \partial_z u_z + A_0^2 C_3 M_4 \partial_z u_z \partial_x^2 u_x + A_0^2 C_3 M_4 \partial_z u_z \partial_x \partial_y u_y 
    \Big]
\end{align} 
and thus we have the following expressions for $\xi, \eta, \phi$ (all other terms are zero)
\begin{align} \label{xixz}
    &\xi_{zxzzz} = \frac{1}{4} C_3 J ; \xi_{zxzzz} = -\frac{1}{4} C_3 J \nonumber \\
    &\xi_{zxzxx} = \xi_{zxxzx} = \xi_{zxxxz} = \frac{1}{6} \left( \frac{J}{2} C_4 + A_0^2 \left( C_3 M_4 - C_4 M_3\right)\right) \nonumber \\
    &\xi_{xzzxx} = \xi_{xzxzx} = \xi_{xzxxz} = -\frac{1}{6} \left( \frac{J}{2} C_4 + A_0^2 \left( C_3 M_4 - C_4 M_3\right)\right) \nonumber \\
    &\eta_{zyxzy} = \eta_{zyxyz} = \eta_{zyyxz} = \eta_{zyyzx} = \eta_{zyzxy} = \eta_{zyzyx} = \frac{1}{12} \left( \frac{J}{2} C_4 + A_0^2 \left( C_3 M_4 - C_4 M_3\right) \right) \nonumber \\
    &\eta_{yzxzy} = \eta_{yzxyz} = \eta_{yzyxz} = \eta_{yzyzx} = \eta_{yzzxy} = \eta_{yzzyx} = -\frac{1}{12} \left( \frac{J}{2} C_4 + A_0^2 \left( C_3 M_4 - C_4 M_3\right) \right) \nonumber \\
    &\eta_{xyzzy} = \eta_{xyzyz} = \eta_{xyyzz} = \frac{1}{6} \left(\frac{J}{2} C_4\right) \nonumber \\
    &\eta_{yxzzy} = \eta_{yxzyz} = \eta_{yxyzz} = - \frac{1}{6} \left(\frac{J}{2} C_4\right)
\end{align} 

We next hope to derive the contribution to the equation of motion from $S_{ax}$. We find
\begin{align} \label{S11}
    \delta S_{11} &= - \int d^3 r \frac{\partial \theta}{\partial y} \xi_{ijkls} \left( \partial_k \delta u_i \partial_l \partial_s u_j + \partial_k u_i \partial_l \partial_s \delta u_j \right) \nonumber \\
    &= - \int d^3 r \frac{\partial \theta}{\partial y} \xi_{ijkls} \left( -  \partial_l \partial_k \partial_s u_j \delta u_i + \partial_l \partial_s \partial_k u_i  \delta u_j \right) \nonumber \\
    &= \int d^3 r \frac{\partial \theta}{\partial y} \left( \xi_{ijkls} - \xi_{jikls} \right)  \partial_k \partial_l \partial_s u_j \delta u_i 
\end{align}
so that the contribution to the equation of motion from the first term is $\frac{\partial \theta}{\partial y} \left( \xi_{ijkls} - \xi_{jikls}\right) \partial_k \partial_l \partial_s u_j $. Similarly, 
\begin{align}\label{S12}
    \delta S_{12} &= - \int d^3 r \frac{\partial \theta }{\partial y} \eta_{ijkl} \left( \partial_k \delta u_i \partial_y \partial_l u_j + \partial_k u_i \partial_y \partial_l \delta u_j \right) \nonumber \\
    &= \int d^3 r \eta_{ijkl} \left(\frac{\partial \theta }{\partial y}    \delta u_i \partial_k \partial_y \partial_l u_j - \partial_y \left( \frac{\partial \theta }{\partial y}  \partial_l \partial_k u_i   \right)\delta u_j \right) \nonumber \\
    &= \int d^3 r \left( \eta_{ijkl} \frac{\partial \theta}{\partial y} \partial_k \partial_y \partial_l u_j - \eta_{jikl} \partial_y \left( \frac{\partial \theta}{\partial y} \partial_k \partial_l u_j \right) \right) \delta u_i
\end{align}
so that the contribution to the equation of motion from the second term is $\eta_{ijkl} \frac{\partial \theta}{\partial y} \partial_k \partial_y \partial_l u_j - \eta_{jikl} \partial_y \left( \frac{\partial \theta}{\partial y} \partial_k \partial_l u_j \right)$;
\begin{align}\label{S13}
    \delta S_{13} &= - \int d^3 r \frac{\partial \theta}{\partial y} \phi_{ijk} \left( \partial_y \delta u_i \partial_y \partial_k u_j + \partial_y u_i \partial_y \partial_k \delta u_j \right)\nonumber \\
    &= - \int d^3 r \phi_{ijk} \left( -\partial_y \left(\frac{\partial \theta}{\partial y} \partial_y \partial_k u_j \right)\delta u_i + \partial_y \left( \frac{\partial \theta}{\partial y} \partial_y \partial_k u_i\right) \delta u_j \right) \nonumber \\
    &= \int d^3 r \left(\phi_{ijk} - \phi_{jik} \right) \partial_y \left( \frac{\partial\theta }{\partial y} \partial_y \partial_k u_j\right) \delta u_i
\end{align}
so that the contribution to the equation of motion from the third term is $\left(\phi_{ijk} - \phi_{jik} \right) \partial_y \left( \frac{\partial\theta }{\partial y} \partial_y \partial_k u_j\right)$. 

Therefore, the full equation of motion can be obtained from $\frac{\delta S}{\delta u_i} = 0$ with $S = S_0 + S_{ax}$ where $S_0$ is given in Eq(\ref{S0general}) and $S_{ax}$ is given by Eqs(\ref{S11general}-\ref{S13general}), which lead to
\begin{align} \label{EOMgeneral}
    \rho \ddot{u_i}  &= \partial_j \left( \lambda_{ijkl} \partial_k u_l \right) + \frac{\partial \theta}{\partial y} \left( \xi_{ijkls} - \xi_{jikls}\right) \partial_k \partial_l \partial_s u_j + \eta_{ijkl} \frac{\partial \theta}{\partial y} \partial_k \partial_y \partial_l u_j - \eta_{jikl} \partial_y \left( \frac{\partial \theta}{\partial y} \partial_k \partial_l u_j \right) \nonumber \\ &+ \left(\phi_{ijk} - \phi_{jik} \right) \partial_y \left( \frac{\partial\theta }{\partial y} \partial_y \partial_k u_j\right).
\end{align}

%\ref{unsrt}
%\ref{sample}

\section{Analytic solution of Axion domain wall modes}\label{AppAnal}

\subsection{Bulk axion domain wall}\label{bulkinf}

We consider periodic boundary conditions along $x$ and $z$ directions and an axion domain at the origin along the y direciton, so the form of the $\theta$ function is given by $\theta(y) = - \frac{\theta_0 }{2\pi^2}\Theta(y)$ where $\Theta(y)$ is the Heaviside step function. We assume the solution to decay exponentially at infinity. We rewrite $ \frac{\theta_0}{2\pi^2} \frac{J}{2}C_3 \equiv r$, $ \frac{\theta_0}{2\pi^2} \frac{J}{2}C_4 \equiv s$ and $\frac{\theta_0}{2\pi^2} A_0^2 (C_3 M_4 - C_4 M_3) \equiv t$ and $\alpha(k_x,k_z) = r k_z^3 -  (s + t) k_x^2 k_z $. We choose the ansatz $\textbf{u}(\textbf{r},t) = \textbf{f}(y)e^{i \textbf{k}_{||}\cdot\textbf{r}_{||}}$ with $\textbf{r}_{||} = (x,z)$ and $\textbf{k}_{||} = (k_x,k_z)$. Then, $H \textbf{f} = \omega^2 \textbf{f}$ where $H = H_0 + H_{ax}$ with

\begin{equation}
    H_0  = \begin{pmatrix} 
     (a+b) k_x^2 + dk_z^2 - b \partial_y^2  & -i k_x a \partial_y  & (c+d) k_x k_z \\
    -i k_x a \partial_y  & b k_x^2 + dk_z ^2 - (a+b) \partial_y^2  & -i k_z (c+d)\partial_y  \\
     (c+d) k_x k_z & -i k_z (c+d)\partial_y & f k_z^2 + d k_x^2 - d \partial_y^2 
    \end{pmatrix},\label{xzH0}
\end{equation}
and
\begin{equation}
    H_{ax}  = \begin{pmatrix} 
     0  &  -\frac{s}{2} k_z^2 \{ \partial_y, \delta(y) \} & -i \delta(y) \alpha \\ 
       \frac{s}{2} k_z^2 \{\partial_y,\delta(y)\} & 0 & +\frac{1}{2} (s + t) k_x k_z  \{\partial_y,\delta(y)\} \\
     i \delta(y) \alpha & -\frac{1}{2} (s + t) k_x k_z \{\partial_y,\delta(y)\} & 0
    \end{pmatrix}, \label{xzHax}
\end{equation}
and the form of the eigenvector $\textbf{f}$ is defined in the main text. Below we will first discuss two simplified situations, in which analytical solutions for the domain wall modes can be obtained.

\subsection{Case I: $\alpha$ term}
In this section, we consider 
the isotropic approximation, where $a = c_l^2 - c_t^2, b = d = c_t^2, c = c_l^2 - 2c_t^2, f = c_l^2$ ,and further assume $c_t = c_l = c_0$, and $r\neq0$, $s=t=0$. Thus, $H_{ph} = H_0 + \delta(y) h_1$ with
\begin{equation} \label{isoxzmatrix}
H_{0} = \begin{pmatrix}
    c_0^2 k^2 - c_0^2 \partial_y^2 & 0 & 0\\
    0  & c_0^2 k^2 - c_0^2 \partial_y^2  & 0 \\
    0 & 0 &  c_0^2 k^2 - c_0^2 \partial_y^2 
    \end{pmatrix},
\end{equation}
and
\begin{equation}
h_1 = \begin{pmatrix}
    0 & 0 & - i \alpha(k_x,k_z) \\
    0  & 0 & 0 \\
    i \alpha(k_x,k_z) & 0 &  0
    \end{pmatrix}. 
\end{equation}

Due to the $\delta$-function in front of $h_1$, we have two regions I and II for $y>0$ and $y<0$, respectively, as shonw in Fig. \ref{fig:S1}a, and choose the ansatz of exponentially localized phonon modes around $y=0$ such that they vanish at infinity i.e. 

\begin{equation}
    \textbf{u} = \begin{cases}
        \sum_{\tau=1}^{3} A_a e^{-\lambda_\tau y} \textbf{u}_0^\tau(y,\lambda_\tau) &  y>0 \\
        \sum_{\tau=1}^{3} B_\tau e^{\xi_\tau y} \textbf{v}_0^\tau(y,\xi_\tau) &  y<0
    \end{cases}
\end{equation}
where $\textbf{u}_0^\tau$ and $\textbf{v}_0^\tau$ are eigenmodes given by $u_i^\tau(y) = e^{-\lambda_\tau y} \delta_{i\tau}$ and $v_i^\tau(y) = e^{\xi_\tau y} \delta_{i\tau}$ with $\lambda_\tau = \xi_\tau = \sqrt{k^2 - \omega^2/c_0^2} > 0 $. By writing down the boundary conditions at $y=0$, we obtain 6 equations for 6 coefficients $A_{1,2,3}$ and $B_{1,2,3}$ given by (for $i=x,y,z$)

\begin{align}
    &\sum_{\tau=1}^3 u_i^\tau(0^+,\lambda_\tau) A_\tau - \sum_{\tau=1}^3 v_i^\tau (0^-,\xi_\tau) B_\tau = 0 \nonumber \\
    &\sum_{\tau=1}^3 \Big[ \sum_j (h_1)_{ij} u_j^\tau(0,\lambda_\tau) + c_0^2 \lambda_\tau u_i^a (0,\lambda_\tau)\Big] A_\tau + \sum_{\tau=1}^3 \Big[ \sum_j (h_1)_{ij} v_j^\tau(0,\xi_\tau) + c_0^2 \xi_\tau v_i^\tau(0,\xi_\tau) \Big] B_\tau =0 \label{case1eq}
\end{align}

Eq(\ref{case1eq}) can be written in a compact form as

\begin{equation}\label{MA1}
    M_{6\times6} \begin{pmatrix}
        A \\
        B
    \end{pmatrix} = 0,
\end{equation}

where \begin{equation}
    M = \begin{pmatrix}
       \mathbb{I}_{3\times 3} & -\mathbb{I}_{3\times 3} \\
       h_1 + c_0^2 \sqrt{k^2 - \omega^2/c_0^2} \mathbb{I}_{3\times 3} & h_1 + c_0^2 \sqrt{k^2 - \omega^2/c_0^2} \mathbb{I}_{3\times 3}
    \end{pmatrix}
\end{equation}
and $A = (A_1, A_2, A_3)^T$ and $B = (B_1, B_2, B_3)^T$. A nontrivial solution exists if the secular equation ($det (M) = 0$) 
\begin{equation}\label{case1dipersion}
    c_0^2 \sqrt{k^2 - \omega^2/c_0^2} \left( c_0^4 \left( k^2 - \omega^2/c_0^2 \right)^2 - \alpha^2 \right) = 0
\end{equation}
is satisfied. Eq(\ref{case1dipersion}) has two solutions, $\omega = c_0 k$ is the bulk mode and $\omega = \left[(c_0 k)^2 - \left( \alpha(k_x,k_z)/c_0 \right)^2 \right]^{1/2}$ is an interfacial mode with a dispersion with the energy lower then the bulk mode, as depicted by the red line in Fig. \ref{fig:S1}b. This gives $\lambda = |\alpha(k_x,k_z)|/c_0^2$. From Eq(\ref{MA1}), we have $A_2 = B_2 = 0$, $A_1 = B_1$ and $A_3 = B_3$, which leads to 
\begin{equation}
    c_0^2 \lambda A_1 - i \alpha A_3 = 0 \implies A_1 = i \frac{\alpha}{c_0^2 \lambda} A_3 = i \frac{\alpha}{c_0^2 |\alpha|/c_0^2} A_3 = i sgn(\alpha) A_3. 
\end{equation}

The interfacial phonon displacement field has the eigenvector form
\begin{equation}
    \textbf{u} = 
        N e^{-\lambda |y|} \begin{pmatrix}
            \mathrm{sgn}(\alpha) \\
            0 \\
            i
        \end{pmatrix} 
\end{equation}
with the normalization factor $N$. The above eigen-function of phonon modes is circularly polarized and the corresponding phonon angular momentum, defined by $l_i = \hbar \textbf{u}_0^{\dagger} M_i \textbf{u}_0$ where $i=x,y,z$ and $(M_i)_{jk} = (-i) \epsilon_{ijk}$ \cite{zhang2014angular,hamada2018phonon}, has non-zero component, $l_y = \mathrm{sgn}(\alpha) \hbar$, which is fully circularly polarized with its helicity depending on the sign of $\alpha$ that in turn depends on the sign of $k_z$ ($\alpha = r k_z^3$).  

\subsection{Case II: only $s \neq 0$}

For \textit{Case II}, we choose $s \neq0$ and $r=s+t=0$. We will show the existence of two interface modes in this limit, which explains Fig. 1d of the main text. Due to the presence of $\{\partial_y, \frac{\partial \theta}{\partial y} \}$ term, we cannot treat $\frac{\partial \theta}{\partial y}$ as $\delta(y)$ analytically. Instead, we consider a linear change in $\theta$. We divide the system into three regions i.e. I ($y>y_0$), II ($-y_0<y<y_0$) and III ($y<-y_0$) as shown in Fig. \ref{fig:S1}c, in which 
\begin{equation}
    \theta(y) = 
    \begin{cases}
        -\theta_0 & y>y_0 \\
        -\theta_0 \frac{y}{y_0} & -y_0<y<y_0 \\
        \theta_0 & y<-y_0
    \end{cases}. 
\end{equation}
The Hamiltonian is found to be block diagonal (when $c+d = 0$) with the $u_z$ part decoupled from $u_{x}$ and $u_y$. We thus focus on the Hamiltonian for $u_{x}$ and $u_y$, which is given by 
\begin{equation} \label{Case2Hsuppl}
H_{ph} = \begin{pmatrix}
    d k_z^2 - b \partial_y^2 & \frac{s}{2} k_z^2 \{\partial_y,\frac{\partial \theta}{\partial y} \} \\
    -\frac{s}{2} k_z^2 \{\partial_y,\frac{\partial \theta}{\partial y} \}  & d k_z^2 - (a+b) \partial_y^2 
    \end{pmatrix},
\end{equation}
and consider the ansatz
\begin{equation}
    \textbf{u} = 
    \begin{cases}
        \sum_{\alpha=1,2} A_\alpha e^{-\lambda_\alpha
        y} \chi_\alpha^{I} & \text{Region I} \\
        \sum_{\alpha=1,2} B_{\alpha',s} e^{i k_{\alpha',s} y} \chi_{\alpha
        }^{II} & \text{Region II} \\
        \sum_{\alpha=1,2} C_\alpha e^{\lambda_\alpha
        y} \chi_\alpha^{III} & \text{Region III}
    \end{cases}. 
\end{equation}

The phonon Hamiltonian for three regions becomes (for $k_x = 0$)
\begin{equation} \label{Case2H}
H_{I} = H_{III} =\begin{pmatrix}
    d k_z^2 - b \lambda^2 & 0 \\
   0  & d k_z^2 - (a+b) \lambda^2
    \end{pmatrix},
\end{equation}

\begin{equation} \label{Case2H}
H_{II} = \begin{pmatrix}
    d k_z^2 + b k^2 & -i \Tilde{\theta} k \\
   i \Tilde{\theta} k  & d k_z^2 + (a+b) k^2
    \end{pmatrix},
\end{equation}
where \begin{equation}
    \Tilde{\theta} = \frac{\theta_0 s}{y_0} k_z^2, \ \lambda_1 = \sqrt{\frac{1}{b}\left( d k_z^2 - \omega^2\right)}, \  \lambda_2 = \sqrt{\frac{1}{a+b}\left( d k_z^2 - \omega^2\right)}; \quad \chi_1^{I} = \begin{pmatrix}
        1 \\ 0
    \end{pmatrix} , \  \chi_2^{I} = \begin{pmatrix}
        0 \\ 1
    \end{pmatrix}
\end{equation}

and 
\begin{equation}
    k_1 = - \frac{1}{\sqrt{2b(a+b)}} \sqrt{(a+2b)(\omega^2 - d k_z^2) + \Tilde{\theta}^2 - \sqrt{a^2(\omega^2 - d k_z^2)^2 + 2 (a+2b)(\omega^2 - d k_z^2)\Tilde{\theta}^2 + \Tilde{\theta}^4}}
\end{equation}

\begin{equation}
    k_2 = - \frac{1}{\sqrt{2b(a+b)}} \sqrt{(a+2b)(\omega^2 - d k_z^2) + \Tilde{\theta}^2 + \sqrt{a^2(\omega^2 - d k_z^2)^2 + 2 (a+2b)(\omega^2 - d k_z^2)\Tilde{\theta}^2 + \Tilde{\theta}^4}}
\end{equation}

\begin{equation}
    \chi_{1} ^{II} = \begin{pmatrix}
        i \left( a k^2 -  \sqrt{a^2 k^4 + 4 k^2 \Tilde{\theta}^2} \right) \\
        2 k \Tilde{\theta}
    \end{pmatrix}, \ \
    \chi_{2} ^{II} = \begin{pmatrix}
        i \left( a k^2 + \sqrt{a^2 k^4 + 4 k^2 \Tilde{\theta}^2} \right) \\
        2 k \Tilde{\theta}
    \end{pmatrix}
\end{equation}

\begin{figure*}
 \includegraphics[width=\textwidth]{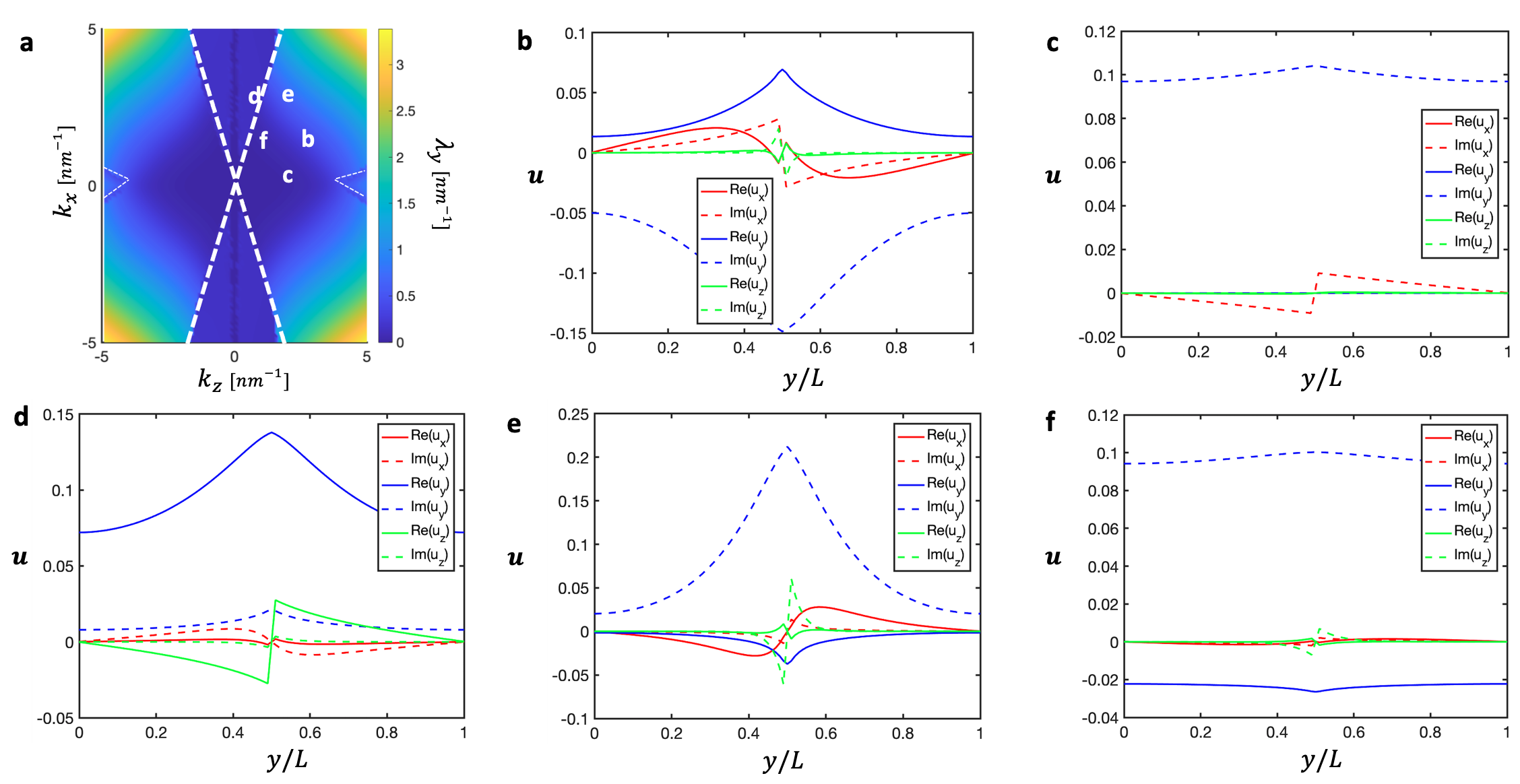}
 \caption{(a) The inverse localization length $\lambda_y$ as a function of $(k_x,k_z)$ and the spatial distribution of the real and imaginary parts of the displacement field of the interface phonon mode for $(k_x,k_z;\lambda_y)$ =  (b) (1,3;0.30) (c) (0,2;0.02)  (d) (2.5,1;0.14)  (e) (2.5,2;0.51)  (f) (1,1;0.02)  [in $nm^{-1}$]. The corresponding We use $r=.8 $; $s=.713 $; $t=.9 $ in units of $10^{-12} m^4/s^2$ and $a=6.75, b=3.17,c=1.01,d=1.93,f=11.02$ in units of $10^6 m^2/s^2$.
 }
 \label{fig:S3}
 \end{figure*}
Since $k$ has to be real, the expression inside the square root must be positive   $ a^2(\omega^2 - d k_z^2)^2 + 2 (a+2b)(\omega^2 - d k_z^2)\Tilde{\theta}^2 + \Tilde{\theta}^4 > 0 $ i.e. $ \omega^2 > d k_z^2 - \frac{\Tilde{\theta}^2}{a+2b+2\sqrt{(a+b)b}}$. For the $\theta$ configuration that is anti-symmetric with respect to $y=0$, both the Hamiltonian $H_0$ and $H_{ax}$ in Eqs(\ref{xzH0},\ref{xzHax}) preserve mirror symmetry $m_y$ and thus all eigenstates should also be the eigenstates of the $m_y$ operator. We thus have two modes, corresponding to even and odd eigenvalues of the $m_y$ operator i.e. $m_y \textbf{u}(y) = \eta \textbf{u} (-y) $  with  $\eta = \pm$ where $m_y$ is defined by  $m_y \begin{pmatrix}
    u_x \\ u_y
\end{pmatrix} = \begin{pmatrix}
    1 & 0 \\ 0 & -1
\end{pmatrix} \begin{pmatrix}
    u_x \\ u_y
\end{pmatrix}.$ 
Due to the $\hat{m}_y$ symmetry, we only need to focus on the interface at $y=y_0$ between Regions I and II. As
\begin{equation}
    \textbf{u}_{II} = \begin{cases}
        \sum_{\alpha=1}^2 \begin{pmatrix}
            B_{\alpha+} \cos(k_\alpha y) \left(a k_\alpha^2 - m(k_\alpha) \right) \\
            B_{\alpha+} \sin(k_\alpha y) (2 k_\alpha \Tilde{\theta}) 
        \end{pmatrix}, & \eta=+ \\
        \sum_{\alpha=1}^2 \begin{pmatrix}
            - B_{\alpha+} \sin(k_\alpha y) \left(a k_\alpha^2 - m(k_\alpha) \right) \\
            B_{\alpha+} \cos(k_\alpha y) (2 k_\alpha \Tilde{\theta}) 
        \end{pmatrix}, & \eta=-
    \end{cases}
\end{equation}
where $m(k_\alpha) = \sqrt{a^2k_\alpha^2 + 4k_\alpha^2 \Tilde{\theta}^2}$, similar to \textit{Case I}, the boundary conditions are given by  
\begin{equation}
    u_{I}(y=y_0) = u_{II}^{\eta=\pm}(y=y_0)
\end{equation}
\begin{equation}
     v_y^{I} u_{I}(y=y_0) = v_y^{II} u_{II}^{\eta= \pm} (y=y_0)
\end{equation}  
\begin{equation}
     \begin{pmatrix}
        -2i b \partial_y & \\
        & -2i(a+b) \partial_y
    \end{pmatrix} \textbf{u}_I(y=y_0) = \begin{pmatrix}
        -2i b \partial_y & - i\Tilde{\theta} \\
        i \Tilde{\theta} 
        & -2i(a+b) \partial_y
    \end{pmatrix} \textbf{u}_{II}^{\eta= \pm}(y=y_0)
\end{equation}

where $v_y^{I,II} = \frac{\partial H_{I,II}}{\partial k_y}$ is the velocity operator. Similar to \textit{Case I}, we rewrite the above as 

\begin{equation}
    G^{\eta=\pm}_{4\times 4} \begin{pmatrix}
        A \\
        B
    \end{pmatrix} = 0
\end{equation}
where $A = (A_1, A_2)^T$ and $B = (B_{1,+},B_{2,+})^T$. Explicitly, 
\begin{equation}\label{G+}
    G^+ = \begin{pmatrix}
        1 & 0 & -\cos(k_1 y_0) g_1 & -\cos(k_2 y_0) g_2 \\
        0 & 1 & -\sin(k_1 y_0) (2 k_1 \Tilde{\theta}) & -\sin(k_2 y_0) (2 k_2 \Tilde{\theta}) \\
        b \lambda_1 & 0 & k_1 \sin(k_1 y_0) (\Tilde{\theta}^2 - b g_1) & k_2 \sin(k_2 y_0) (\Tilde{\theta}^2 - b g_2) \\
        0 & (a+b) \lambda_2 & \Tilde{\theta} \cos(k_1 y_0) \left( 2(a+b) k_1^2 - g_1/2\right) & \Tilde{\theta} \cos(k_2 y_0) \left( 2(a+b) k_2^2 - g_2/2\right)
    \end{pmatrix}
\end{equation}
and
\begin{equation}\label{G-}
    G^- = \begin{pmatrix}
        1 & 0 & \sin(k_1 y_0) g_1 & \sin(k_2 y_0) g_2 \\
        0 & 1 & -\cos(k_1 y_0) (2 k_1 \Tilde{\theta}) & -\cos(k_2 y_0) (2 k_2 \Tilde{\theta}) \\
        b \lambda_1 & 0 & k_1 \cos(k_1 y_0) (\Tilde{\theta}^2 - b g_1) & k_2 \cos(k_2 y_0) (\Tilde{\theta}^2 - b g_2) \\
        0 & (a+b) \lambda_2 & -\Tilde{\theta} \sin(k_1 y_0) \left( 2(a+b) k_1^2 - g_1/2\right) & -\Tilde{\theta} \sin(k_2 y_0) \left( 2(a+b) k_2^2 - g_2/2\right)
    \end{pmatrix}
\end{equation}
where $g_i = ak_i^2 - \sqrt{a^2 k_i^4 + 4 \Tilde{\theta}^2 k_i^2}$ for $i=1,2$. We solve the secular equation $det (G^{\pm}) =0$ numerically and find the dispersion of two interface modes, depicted by the red and dark green lines in Fig. \ref{fig:S1}d. 

\begin{figure*}
 \includegraphics[width=\textwidth]{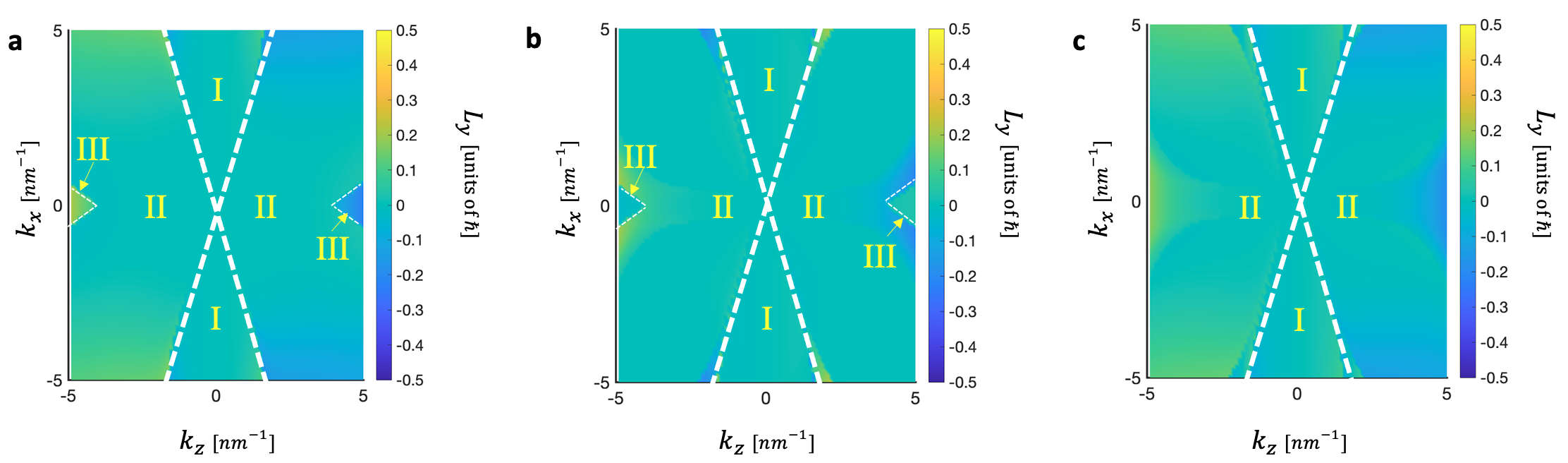}
 \caption{(a) The out of plane ($y$ component) phonon angular momentum of the lowest mode as a function of $(k_x,k_z)$ (b) The out of plane (y component) phonon angular momentum of the second lowest mode as a function of $(k_x,k_z)$ (c) The out of plane (y component) phonon angular momentum of the lowest two modes added together as a function of $(k_x,k_z)$. We use $r=.8 $; $s=.713 $; $t=.9 $ in units of $10^{-12} m^4/s^2$  and $a=6.75, b=3.17,c=1.01,d=1.93,f=11.02$ in units of $10^6 m^2/s^2$
}
 \label{fig:S2}
 \end{figure*}

\section{Numerical method} \label{AppNumerics}
In this section, we outline the numerical methods for solving the lattice model version of Eigen-equation (4) in the main text with an axion domain wall perpendicular to the $x-z$ plane. We consider a lattice regularization along $y$ by dividing the $y$ axis into $N+1$ discrete sites with periodic boundaries at $i=1$ and $i=N+1$.

\subsection{$H_0$ term}
We define the components of $\Vec{u}$ as $u_i^x = \varphi_{3i-2},u_i^y = \varphi_{3i-1}, u_i^z = \varphi_{3i}$ where $i=1,2,...,N+1$. We start with $-\rho \omega^2 u_i = \partial_j (\lambda_{ijkl} \partial_k u_l)$ and define $L_{in,lm}$ such that $-\rho \omega^2 u_i(y_n) = L_{in,lm} u_l(y_m)$ with $L = L^{(0)}+L^{(1)}+L^{(2)}$. Here, $u_l(y_m) = \varphi_{3(m-1)+l}$

\begin{align}\label{NumH0}
    L_{in,lm}^{(0)} &= - k_j k_k \lambda_{ijkl} \delta_{mn} \nonumber \\
    L_{in,ln+1}^{(1)} &= \frac{i k_k}{2h} \left( \lambda_{iykl}(y_{n+1}) + \lambda_{lyki}(y_n) \right) \nonumber \\
    L_{l n+1,in}^{(1)} &= L_{in,ln+1}^{(1)^*} \nonumber \\
    L_{in,ln+1}^{(2)} &=\frac{1}{2h^2} \left( \lambda_{iyyl}(y_{n+1}) + \lambda_{iyyl}(y_n) \right) \nonumber \\
    L_{ln+1,in}^{(2)} &= L_{in,ln+1}^{(2)^*} \nonumber \\
    L_{in,ln}^{(2)} &= -\frac{1}{2h^2} \left( \lambda_{iyyl}(y_{n+1}) + \lambda_{iyyl}(y_{n-1}) + 2 \lambda_{iyyl}(y_n)  \right)
\end{align}

\begin{figure*}
 \includegraphics[width=\textwidth]{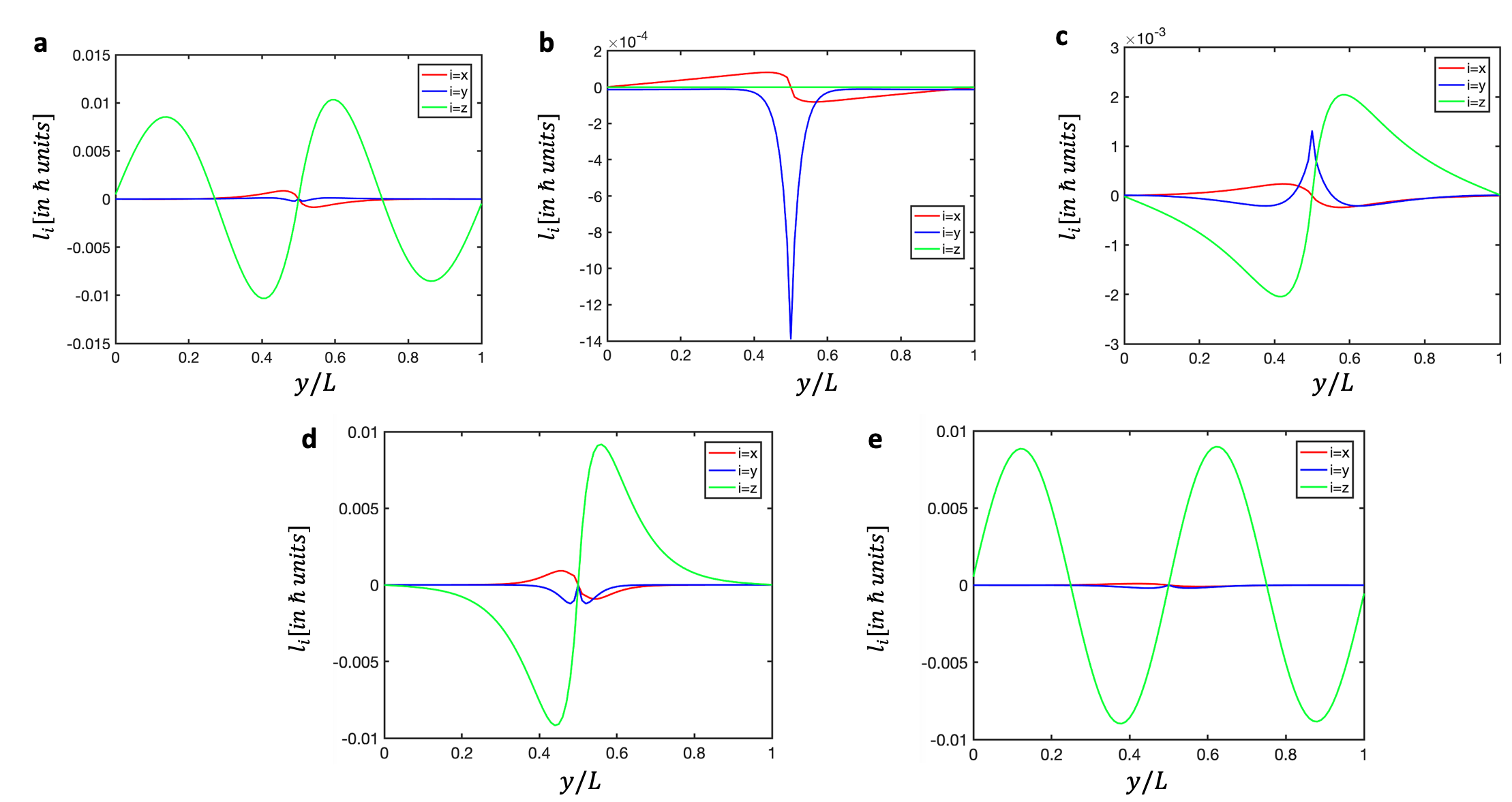}
 \caption{The spatial distribution of phonon angular momentum of the interface phonon modes at $(k_x,k_z)$ =  (a) (1,3) (b) (0,2) (c) (2.5,1) (d) (2.5,2) (e) (1,1) [in $nm^{-1}$]. We use $r=.8 $; $s=.713 $; $t=.9 $ in units of $10^{-12} m^4/s^2$ and $a=6.75, b=3.17,c=1.01,d=1.93,f=11.02$ in units of $10^6 m^2/s^2$
 }
 \label{fig:S4}
 \end{figure*}

\subsection{$H_1$ term}

In the presence of the axion term, the total Hamiltonian is $H=H_0 + H_1$. Using Eqs(\ref{S11},\ref{S12}, \ref{S13}), we define the following operators using Einstein's summation convention

\begin{align} \label{NumH1}
    \big[L_{11}\big]_{in,jm} &= i \frac{\partial \theta}{\partial y} (y_m) \left( \xi_{ijkls} - \xi_{jikls} \right) k_k k_l k_s \delta_{nm} \nonumber \\
    \big[L_{12}\big]_{in,jn+1} &= \left(\eta_{ijkl} \frac{\partial \theta}{\partial y}(y_n) - \eta_{jikl} \frac{\partial \theta}{\partial y} (y_{n+1}) \right) \frac{k_k k_l}{2h} \nonumber \\
    \big[L_{12}\big]_{jn+1,in} &= \big[L_{11}\big]_{in,jn+1}^* \nonumber \\
    \big[L_{13}\big]_{in,jn+1} &= - \frac{i k_k}{2 h^2} \left( \phi_{ijk} - \phi_{jik} \right) \left( \frac{\partial \theta}{\partial y} (y_n) + \frac{\partial \theta}{\partial y} (y_{n+1}) \right) \nonumber \\
    \big[L_{13}\big]_{jn+1,in} &= \big[L_{11}\big]_{in,jn+1}^* \nonumber \\
    \big[L_{13}\big]_{in,jn} &= \frac{i k_k}{2 h^2} \left( \phi_{ijk} - \phi_{jik} \right) \left(  \frac{\partial \theta}{\partial y} (y_{n+1}) + \frac{\partial \theta}{\partial y} (y_{n-1}) +2 \frac{\partial \theta}{\partial y} (y_{n}) \right)  
\end{align}

We consider the following axion domain wall (the discrete version fo the step function)

\begin{equation}
  \frac{\partial \theta}{\partial y}(y_n)=\begin{cases}
    -\theta_0/h, & \text{if $n=N/2$}.\\
    0, & \text{otherwise}.
  \end{cases}
\end{equation}

\begin{figure*}
 \includegraphics[width=\textwidth]{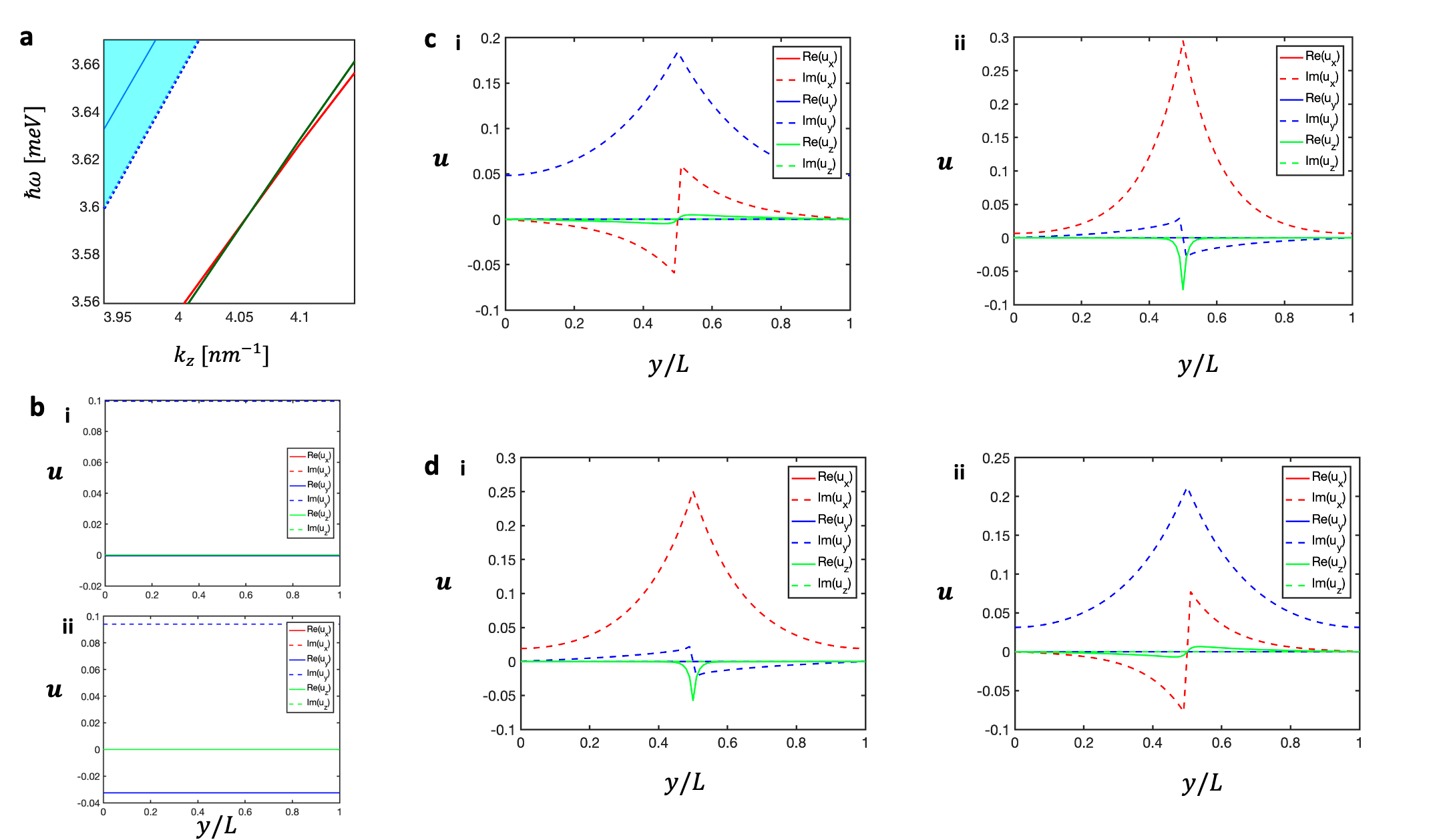}
 \caption{(a) Band crossing between two interface phonon modes at around $k_z=4.05 nm^{-1}$ and $k_x=0$. This is also the inset in Fig. 1d of the main text. (b) The spatial distribution of the real and imaginary parts of the displacement field of the lowest bulk phonon mode at (i) $k_z= 3.9 nm^{-1}$ and (ii) $k_z= 4.2 nm^{-1}$ (c) The spatial distribution of the real and imaginary parts of the displacement field of the lowest interface phonon mode at (i) $k_z= 3.9 nm^{-1}$ and (ii) $k_z= 4.2 nm^{-1}$ (d) The spatial distribution of the real and imaginary parts of the displacement field of the second lowest interface phonon mode at (i) $k_z= 3.9 nm^{-1}$ and (ii) $k_z= 4.2 nm^{-1}$ }
 \label{fig:S5}
 \end{figure*}

\subsection{Numerical Results}
Here we will discuss our numerical results in Fig. \ref{fig:S3}-\ref{fig:S5}, to support our discussion of the interfacial modes in the main text. Fig. \ref{fig:S3} shows the spatial profile of the real and imaginary parts of the displacement field for the interface phonon modes at five different momenta, which all reveal exponentially decaying behaviors away from the axion domain wall located at $y=L/2$. For small $|\textbf{k}_{\parallel}|$, the spatial distribution is uniform, but gets more localized with increasing $|\textbf{k}_{\parallel}|$. 

Fig. \ref{fig:S2} and Fig. \ref{fig:S4} provide information of the angular momentum of interfacial model modes. 
We find the interfacial phonon mode has a nonzero angular momentum component along the $y$ direction $l_y$, and thus we show $l_y$ as a function of ${\bf k}_{\parallel}$ in Fig. \ref{fig:S2}. For most of $\textbf{k}_{\parallel}$ plane, the dominant phonon angular momentum comes from the lowest energy phonon mode (Fig. \ref{fig:S2}a), whereas the phonon angular momentum of the second lowest energy phonon mode is vanishingly small (Fig. \ref{fig:S2}b). However, we notice a rapid change of the phonon angular momenta for both the lowest and second lowest modes between the Region II and Region III, which is due to the crossing in energy between these two low energy phonon modes. The band crossing between the two interfacial modes is shown in Fig. \ref{fig:S5}a which occurs around $k_z=4.05 nm^{-1}$ and $k_x=0$. The spatial distribution of the real and imaginary parts of the displacement field of the two interfacial modes is shown in Fig. \ref{fig:S5}c,d to conclusively identify their band crossing. The lowest bulk mode \ref{fig:S5}b shows no change. We also plot the spatial variation of $l_i$ with $i=x,y,z$ as a function of $y$ for various values of $\textbf{k}_{\parallel}$ in Fig. \ref{fig:S4}. The $l_x$ and $l_z$ components of the phonon angular momentum are anti-symmetric with respect to the axion domain wall at $y=L/2$, whereas the $l_y$ component is symmetric. Thus, $l_x$ and $l_z$ vanish after summing over the y-locations, while $l_y$ is still non-zero. Further, as $\textbf{k}_{\parallel}$ increases, the magnitude of $\sum_y l_y(y)$ increases.

\section{Surface phonon modes}\label{secsurf}
We have discussed the existence of interface phonon modes. It is well known that surface acoustic wave  can exist in an elastic material \cite{landau1986theory,kress1991surface}. In this section, we will study the influence of the axion term on the surface acoustic wave. 

\subsection{$xz$ plane configuration}

We consider a thin film model with the top ($y=y_1$) and bottom ($y=y_2$)surfaces parallel to the $x-z$ plane. In the absence of the axion term, the equation of motion is given by

\begin{equation}
    \rho \ddot{u}_i = \partial_{j}\sigma_{ij} = \partial_j \frac{\partial F_0}{\partial u_{ij}}= \partial_j \left( \lambda_{ijkl} \partial_k  u_l \right)
\end{equation}

The elastic waves vanish at the bottom surface ($y=y_1$) and the top surface ($y=y_2$) is considered a stress free surface i.e. $u_x(y=y_1) = u_y(y=y_1)=u_z(y=y_1)=0$ and $\sigma_{xy}(y=y_2) = \sigma_{yz}(y=y_2)=\sigma_{yy}(y=y_2)=0$ where the stress tensors are

\begin{align}
    \sigma_{xy} = \frac{\partial F}{\partial u_{xy}} &= 2 b u_{xy} \label{xzbcxy}\\
    \sigma_{yz} = \frac{\partial F}{\partial u_{yz}} &= 2 d u_{yz} \label{xzbcyz}\\
    \sigma_{yy} =\frac{\partial F}{\partial u_{yy}} &= (a+b) u_{yy} + (a-b) u_{xx} + c u_{zz} \label{xzbcyy}
\end{align}

 It is possible to grow the sample on a substrate such that the top surface obeys free boundary condition whereas the bottom surface obeys open boundary conditions when the substrate density is much greater than the sample density. We impose periodic boundary conditions along $x$ and $z$ directions. We consider the wave ansatz of the form $u_i  = f_i(y) e^{i\left( k_x x + k_z z - \omega t \right)}$. Therefore, $k_x$ and $k_z$ are good quantum numbers so that we can replace $\partial_x \rightarrow ik_x$ and $\partial_z \rightarrow ik_z$. After acting on the Ansatz, we simplify the equation of motion to the form $H_0 \textbf{f} = \rho w^2 \textbf{f}$ where 

\begin{equation}
    H_0  = \begin{pmatrix} 
     (a+b) k_x^2 + dk_z^2 - b \partial_y^2 -(\partial_y b)\partial_y & -i k_x( a \partial_y + (\partial_yb) ) & (c+d) k_x k_z \\
    -i k_x( a \partial_y + (\partial_yb) ) & b k_x^2 + dk_z ^2 - (a+b) \partial_y^2 - \partial_y(a+b) \partial_y & -i k_z ((c+d)\partial_y + (\partial_yc)) \partial_y \\
     (c+d) k_x k_z & -i k_z ((c+d)\partial_y + (\partial_yc)) \partial_y & f k_z^2 + d k_x^2 - d \partial_y^2 - (\partial_y d) \partial_y
    \end{pmatrix}
\end{equation}

We denote the effective action for the axion term to be of the form 

\begin{equation}
    S_{ax} = - \int d^3 r \frac{\partial \theta (y)}{\partial y} \Big[ \xi_{ijkls} \partial_k u_i \partial_l \partial_s u_j + \eta_{ijkl} \partial_k u_i \partial_y \partial_l u_j + \phi_{ijk} \partial_y u_i \partial_y \partial_k u_j \Big]  
\end{equation}
with $i,j = x,y,z$, $k,l,s = x,z$ and $\xi, \eta, \phi$ are given in Eq(\ref{xixz}). Specifically, the axion contribution for the above thin film model is given by Eq(\ref{S1xz}), 
\begin{align}
    S_{ax} &= \int d^3 r \frac{\partial \theta}{\partial y} \Big[ -\frac{J}{2} C_3 \partial_z u_z \partial_z^2 u_x  - \frac{J}{2} C_4 \partial_x u_x \partial_x \partial_z u_z - \frac{J}{2} C_4 \partial_y u_y \partial_z^2 u_x - \frac{J}{2} C_4 \partial_y u_y \partial_x \partial_z u_z \nonumber \\ &+ A_0^2 C_4 M_3 \partial_x u_x \partial_x \partial_z u_z + A_0^2 C_4 M_3 \partial_y u_y \partial_x \partial_z u_z + A_0^2 C_3 M_4 \partial_z u_z \partial_x^2 u_x + A_0^2 C_3 M_4 \partial_z u_z \partial_x \partial_y u_y 
    \Big].
\end{align}

The equation of motion is given by Eq(\ref{EOMgeneral}) and Eq(\ref{xixz})i.e. $H \Vec{f} = \rho \omega^2 \Vec{f}$ , where $H = H_0 + H_1$ with $H_1 = - \left( \partial_y \theta  h_1 + \frac{1}{2} \partial_y^2 \theta h_2 \right) $. We have used the redefined parameters $r,s,t,\alpha$ from Sec. \ref{bulkinf},

\begin{equation}\label{surfh1h2}
    h_1  = \begin{pmatrix} 
     0  & - s k_z^2 \partial_y & - i \alpha \\ s k_z^2 \partial_y & 0 & (s+t) k_x k_z \partial_y \\ 
    i \alpha & - (s+t) k_x k_z \partial_y & 0 
    \end{pmatrix}, \ \ 
    h_2  = \begin{pmatrix} 
     0  & - s k_z^2  & 0\\ s k_z^2  & 0 & (s+t) k_x k_z  \\ 
    0 & - (s+t) k_x k_z & 0 
    \end{pmatrix}
\end{equation}

\begin{figure*}
 \includegraphics[width=\textwidth]{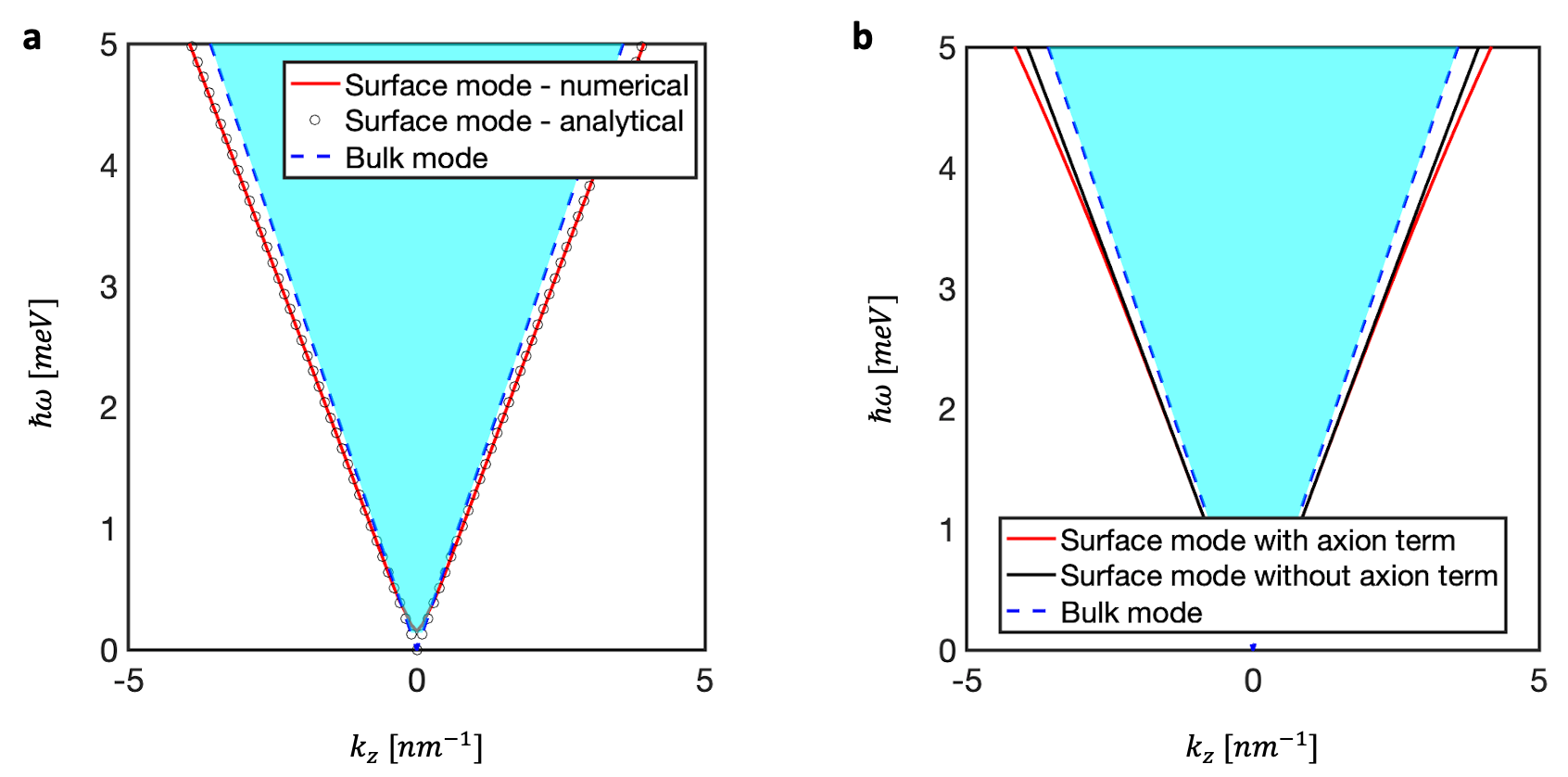}
 \caption{(a) The existence of a surface phonon mode even in the absence of the axion term is shown, where the numerical result is depicted by the solid red line and the analytical result is depicted by black circles.  (b) In the presence of the axion term, the energy of surface modes is lowered and is no longer linear in $\textbf{k}$. We use $r=.8 $; $s=.713 $; $t=.9 $ in units of $10^{-12} m^4/s^2$. For both the plots we consider the $k_x = k_z$ line and the isotropic approximation with $c_t = 1.5$ and $c_l =1.5$ in units of $10^6 m^2/s^2$.}
 \label{fig:S6}
 \end{figure*}

\subsection{Isotropic approximation for the bulk action without the Axion term}\label{sec:Slandau}

The isotropic free energy is given by

\begin{equation}
    F = \frac{\rho}{2} \left( c_l^2 -2 c_t^2 \right) u_{ii}^2 + \rho c_t^2 u_{ik}^2
\end{equation}

The hexagonal crystal free energy given in Eq(\ref{5const}) boils down to the above isotropic limit when $a = \rho \left( c_l^2 - c_t^2 \right), b = d = \rho c_t^2, c= \rho \left( c_l^2 - 2 c_t^2\right), f = \rho c_l^2$ The equation of motion is thus:

\begin{equation} \label{wave0}
    \ddot{\pmb{u}} = c_t^2 \nabla^2 \pmb{u} + \left( c_l^2 - c_t^2\right) \pmb{\nabla} \left( \pmb{\nabla}\cdot \pmb{u}\right)
\end{equation}
Substituting the elastic wave ansatz form 
%Plugging this 
into Eq. (\ref{wave0}), we have the eigen equation $H_0\textbf{f} = \omega^2 \textbf{f}$ with 
\begin{equation} \label{isoxzmatrix}
H_0 = \begin{pmatrix}
    c_l^2 k_x^2 + c_t^2 k_z^2 - c_t^2 \partial_y^2 - (\partial_y c_t^2)\partial_y& - i k_x \left( c_l^2 - c_t^2 + (\partial_y c_t^2)\right) \partial_y & k_x k_z \left( c_l^2 - c_t^2\right) \\
    -i k_x \left( c_l^2 - c_t^2 + \partial_y (c_l^2 - 2 c_t^2)\right)\partial_y  & c_t^2 k_x^2 + c_t^2 k_z^2 - c_l^2 \partial_y^2 - (\partial_y c_l^2) \partial_y & - i k_z \left( c_l^2 - c_t^2 + \partial_y(c_l^2 - 2c_t^2)\right) \partial_y \\
    k_x k_z \left( c_l^2 - c_t^2\right) & - i k_z \left( c_l^2 - c_t^2 + \partial_y c_t^2\right) \partial_y &  c_t^2 k_x^2 + c_l^2 k_z^2 - c_t^2 \partial_y^2 -(\partial_y c_t^2) \partial_y
    \end{pmatrix}. 
\end{equation}

The stress tensors become

\begin{align}
    \sigma_{xy} = \frac{\partial F}{\partial u_{xy}} &= 2 \rho c_t^2 u_{xy} \label{isoxzbcxy}\\
    \sigma_{yz} = \frac{\partial F}{\partial u_{yz}} &= 2 \rho c_t^2 u_{yz} \label{isoxzbcyz}\\
    \sigma_{yy} =\frac{\partial F}{\partial u_{yy}} &= \rho c_l^2 u_{yy} + \rho (c_l^2 -2 c_t^2)  \left( u_{xx} +  u_{zz} \right) \label{isoxzbcyy}
\end{align}

The surface elastic wave solution of the isotropic model can be solved analytically (in the absence of the axion term), as presented in 
%In this section, we will follow the derivation presented in 
Refs. \cite{landau1986theory,kress1991surface}. We start out with the equation of motion in Eq. \ref{wave0} and decompose the displacement field into the longitudinal and transverse components, $\pmb{u_t} + \pmb{u_l}$ 
$\pmb{u}=\pmb{u_t} + \pmb{u_l}$, with $\pmb{\nabla}\cdot\pmb{u_t} = 0$ and $\pmb{\nabla} \times \pmb{u_l} = 0$. Since we consider a semi-infinite system, we expect a surface wave solution of the form $u_{t,i} = \alpha_i e^{ik_x x+ i k_z z- i\omega t + \kappa_t y}$ and $u_{l,i} = \beta_i e^{ik_x x+ i k_z z - i\omega t + \kappa_l y}$ with $\kappa_{t,l} = \sqrt{k^2 - \omega^2/c_{t,l}^2}$. The surface wave dispersion is of the form 
\begin{equation} \label{eq:SurfaceAscoustic}
\omega = \xi c_t k
\end{equation}
with  $k = \sqrt{k_x^2 + k_z^2}$, $\xi<1$, which is determined from 
\begin{eqnarray}\xi^6 - 8 \xi^4 + 8 \left( 3 - 2 c_t^2/c_l^2 \right) \xi^2 -16 \left(1- c_t^2/c_l^2 \right) =0.
\end{eqnarray} 
The surface wave solution is

\begin{equation} \label{eigenvectorlandau}
    \pmb{u} = N \begin{pmatrix}
k_x \Big[ e^{\kappa_t y} - \frac{1}{2 \kappa_t \kappa_l} \left( \kappa_t^2 + k^2\right) e^{\kappa_l y} \Big]\\
- \frac{i}{\kappa_t} \Big[ k^2 e^{\kappa_t y} - \left(\kappa_t^2 + k^2 \right) e^{\kappa_l y} \Big]\\
k_z \Big[ e^{\kappa_t y} - \frac{1}{2 \kappa_t \kappa_l} \left( \kappa_t^2 + k^2\right) e^{\kappa_l y} \Big] 
\end{pmatrix} e^{i k_x x + ik_z z -i \omega t}
\end{equation}
with the normalization factor $N$. 

\begin{figure*}
 \includegraphics[width=\textwidth]{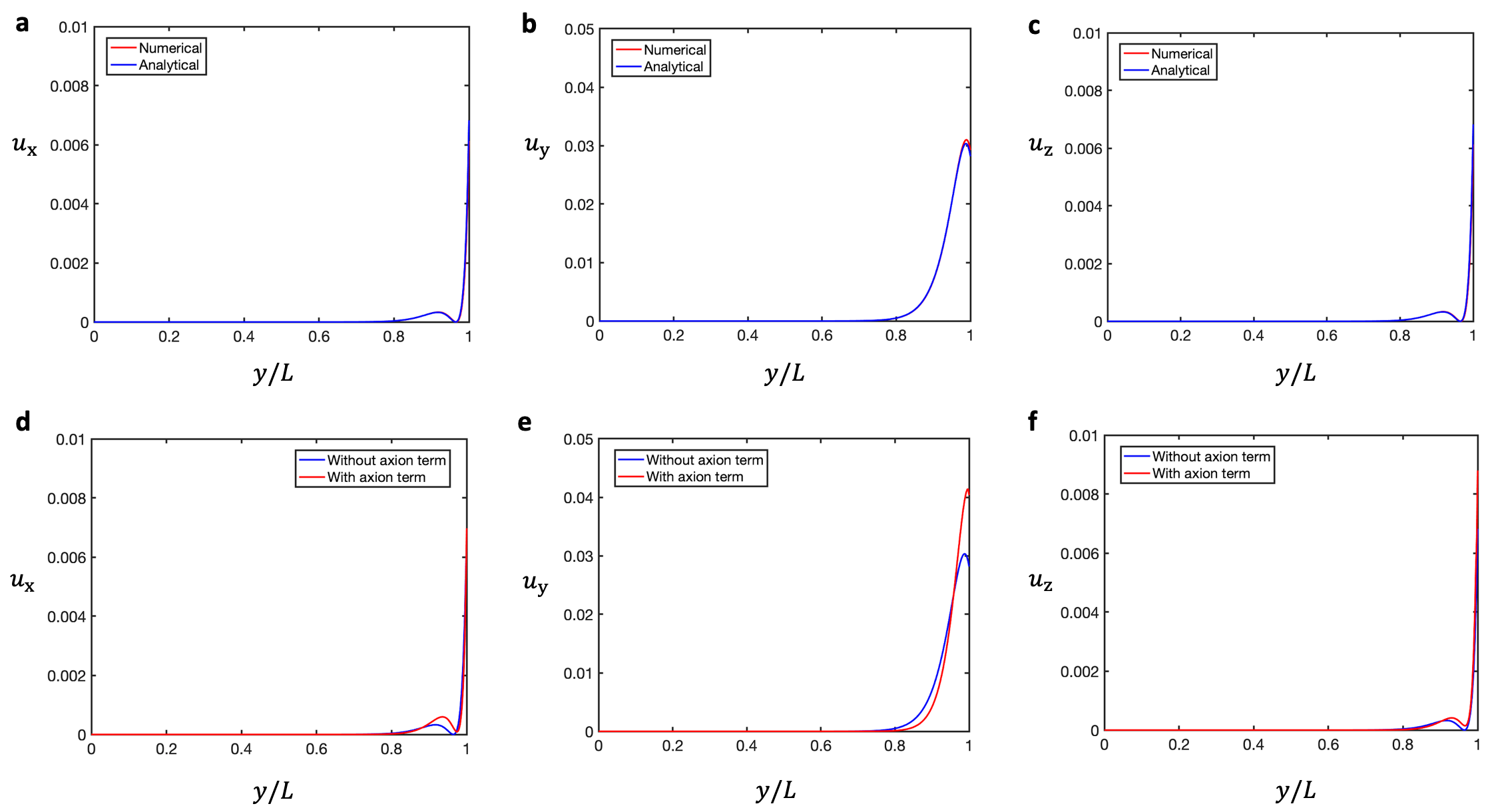}
 \caption{(a) The components of the surface phonon mode displacement field are shown as (a) $u_x$ (b) $u_y$  (c) $u_z$ in the absence of the axion term. In the presence of the axion term, the modified components of surface phonon mode displacement field are shown as (d) $u_x$ (e) $u_y$  (f) $u_z$. We use $r=.8 $; $s=.713 $; $t=.9 $ in units of $10^{-12} m^4/s^2$. For both the plots we consider $k_x = k_z = 2.5 nm^{-1}$ and the isotropic approximation with $c_t = 1.5$ and $c_l =1.5$ in units of $10^6 m^2/s^2$.}
 \label{fig:S7}
 \end{figure*}

\subsection{Numerical method}

We follow the numerical calculation along the lines of Eqs(\ref{NumH0},\ref{NumH1}) in the isotropic approximation with 

\begin{equation}\label{surftheta}
  \frac{\partial \theta}{\partial y}(y_n)=\begin{cases}
    -\theta_0/h, & \text{if $n=N+1$}.\\
    0, & \text{otherwise}.
  \end{cases}
\end{equation}

Stress free boundary conditions in Eqs(\ref{isoxzbcxy}-\ref{isoxzbcyy}) at site (boundary)$N+1$ can be written generally as

\begin{align}\label{surfbc}
     &\sigma_{iy}(y_{N+1})=0 \implies \sum_{ql} \lambda_{iyql} \partial_q u_{l}(y_{N+1}) = 0 \nonumber \\ \implies &\sum_{q=x,z;l} i k_q \frac{\lambda_{iyql}(y_{N+1})+\lambda_{lyqi}(y_{N+2})}{2} u_l(y_{N+1}) + \sum_l \frac{\lambda_{iyyl}(y_{N+2})+ \lambda_{iyyl}(y_{N+2})}{2} \frac{u_l(y_{N+2})-u_l(y_{N+1})}{h}=0 \nonumber \\
    \implies &\sum_{q=x,z;l} - \frac{i k_q}{2h} \Big[ \lambda_{iyql}(y_{N+1})+\lambda_{lyqi}(y_{N+2})\Big] u_l(y_{N+1}) + \sum_l \frac{1}{2h^2} \Big[] \lambda_{iyyl}(y_{N+2}) + \lambda_{iyyl} (y_{N+1})\Big] u_l (y_{N+1}) \nonumber \\ &- \sum_l \frac{1}{2h^2} \Big[] \lambda_{iyyl}(y_{N+2}) + \lambda_{iyyl} (y_{N+1}) \Big] u_l (y_{N+2}) = 0
\end{align}
    
Here, we choose $\lambda_{lyqi}(y_{N+2}) = \lambda_{iyql}(y_{N+1})$ and $\lambda_{iyyl}(y_{N+2}) = \lambda_{lyyi}(y_{N+2})=\lambda_{iyyl}(y_{N+1})$. We make $\partial_q \rightarrow i k_q$ when $q=x,z$. Finally, we solve Eqs(\ref{NumH0},\ref{NumH1},\ref{surftheta},\ref{surfbc}).

\begin{figure*}
 \includegraphics[width=\textwidth]{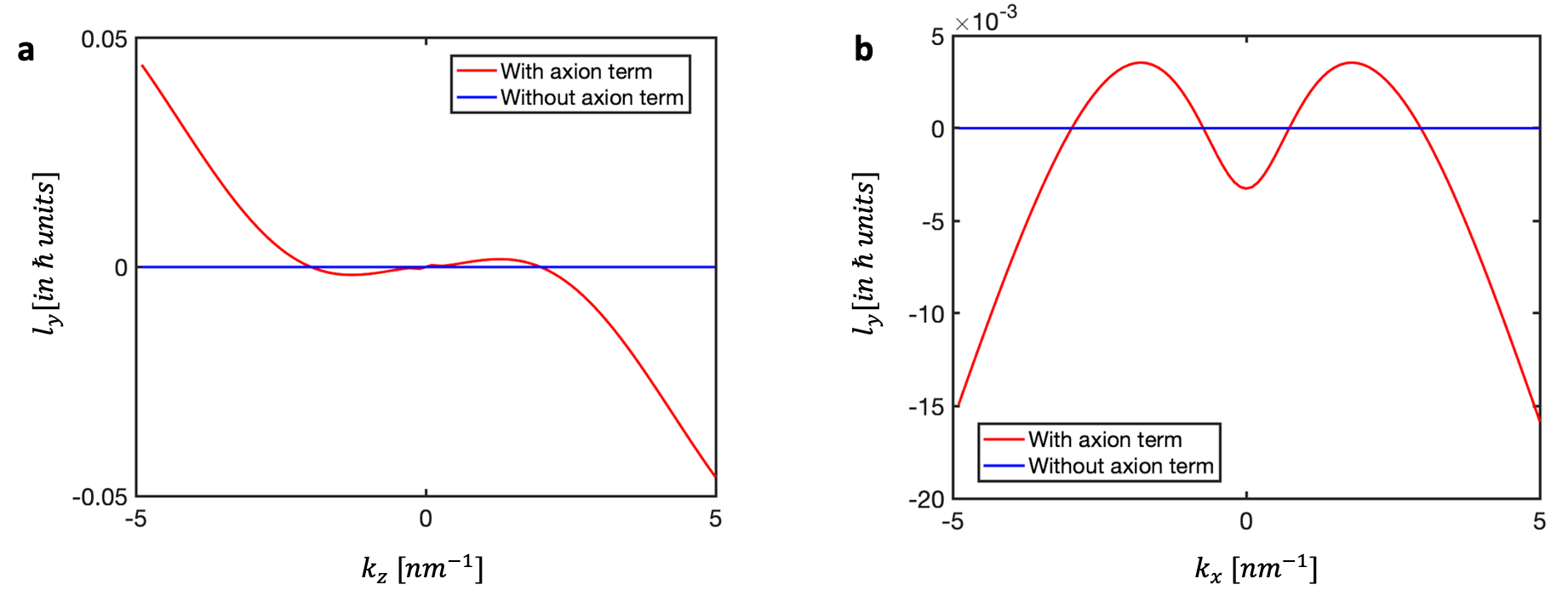}
 \caption{(a) The out-of-plane ($y$-direction) angular momentum of the surface phonon mode $l_y$ as a function of $k_z$ along the $k_x = k_z$ line (b) The out-of-plane ($y$-direction) angular momentum of the surface phonon mode $l_y$ as a function of $k_x$ for $k_z = 1 nm^{-1}$. For both the plots, we use $r=.8 $; $s=.713 $; $t=.9 $ in units of $10^{-12} m^4/s^2$ and the isotropic approximation with $c_t = 1.5$ and $c_l =1.5$ in units of $10^6 m^2/s^2$.}
 \label{fig:S8}
 \end{figure*}

\subsection{Numerical Results}
In this section, we study the influence of the axion term on the surface acoustic phonon modes and the main numerical results are shown in Fig. \ref{fig:S6}-\ref{fig:S8}.  
In the absence of the axion term, from Sec.\ref{sec:Slandau}, there exists a surface acoustic phonon mode with a linear frequency dispersion below the bulk bands, as shown in Fig. \ref{fig:S6}a, where the energy calculated numerically matches pretty well with the analytical expression (\ref{eq:SurfaceAscoustic}) in Sec.\ref{sec:Slandau}. The axion term further lowers the frequency of surface phonon modes, as shown in Fig. \ref{fig:S6}b. In the absence of the axion term, all the three components of the surface phonon displacement field $\textbf{u}$ are localized around the surface $y=L$ and exponentially decay away from the surface, as shown in Fig. \ref{fig:S7}(a-c). For these plots, along $k_x=k_z$, $u_x$ and $u_z$ follow the same decaying behavior, consistent with Eq.(\ref{eigenvectorlandau}). In the presence of the axion term, all the three components of the surface phonon displacement field $\textbf{u}$ are further localized around the surface $y=L$, as shown in Fig. \ref{fig:S7}(d-f). Due to the anistropy in the axion term, $u_x$ and $u_z$ follow different behaviors. The major qualitative change induced by the axion term is the appearance of a non-zero out-of-plane ($y$-direction) angular momentum ($l_y$) of the surface phonon mode, as demonstrated in Fig. \ref{fig:S8}a and b. We observe that $l_y(k_x,k_z) = - l_y(-k_x,-k_z)$, from Fig. \ref{fig:S8}a, in which we plot $l_y$ along the $k_x=k_z$ line, and $l_y(k_x,k_z) = l_y(-k_x,k_z)$, from Fig. \ref{fig:S8}b, in which $k_z$ is a fixed number. The former can be understood from time reversal $\hat{T}$, while the latter is a consequence of the combined effect of $\hat{T}$ and z-directional mirror $\hat{m}_z$. The total phonon angular momentum, summed over all momenta $k_x$ and $k_z$ is zero. Therefore, in order to probe the surface phonon mode via its angular momentum properties, a thermal gradient is required to bring the system away from equilibrium, which has been discussed in details in Sec. \ref{appangmon}.

\subsection{Hermiticity of the Hamiltonian}

In this section, we discuss the hermiticity of our Hamiltonian. In particular, we show that for a free boundary condition, the continuum model Hamiltonians in Eqs(\ref{xzH0},\ref{surfh1h2}) are hermitian.

\subsubsection{Hermiticity of Eq(\ref{xzH0})}

Consider $\langle \pmb{v}|H_0|\pmb{u} \rangle$ where $H_0$ is given by Eq(\ref{xzH0}), $\pmb{u} = \left(u_x (y), u_y(y), u_z(y) \right)^{T}$ and $\pmb{v} = \left(v_x (y), v_y(y), u_z(y) \right)^{T}$. Here, the domain of $H_0$, $D(H_0) = \left\{ \pmb{u} \in H^2(0,L) : u_x(0) = u_y(0) = u_z(0) = 0, \sigma_{xy}(L) = \sigma_{yz}(L) = \sigma_{yy}(L) = 0 \right\} $. The $\sigma$'s are given by Eqs(\ref{xzbcxy}-\ref{xzbcyy}).

\begin{align}
    \langle \pmb{v}| H_0 | \pmb{u}\rangle &= \int_0^L dy v_x^* \Big[ \left((a+b) k_x^2 + d k_z^2 - \rho \omega^2 - b \partial_y^2 \right) u_x - i a k_x \partial_y u_y + (c+d) k_x k_z u_z \Big] \nonumber \\ &+ \int_0^L dy v_y^* \Big[ -i a k_x \partial_y u_x + \left( b k_x^2 + d k_z^2 - \rho \omega^2 - (a+b) \partial_y^2\right) u_y - i (c+d) k_z \partial_y u_z \Big] \nonumber \\ &+ \int_0^L dy v_z^* \Big[ (c+d) k_x k_z u_x  - i(c+d) k_z \partial_y u_y + 
    \left( f k_z^2 + d k_x^2 - \rho \omega^2 - d \partial_y^2 \right) u_z\Big] \nonumber \\ &= \overline{\langle \pmb{u}| H_0^\dagger | \pmb{v}\rangle} + \chi 
\end{align}
with \begin{align}
    \chi &= - b v_x^{*} \left(\partial_y u_x \right)|_0^L + b \left(\partial_y u_x^{*} \right)|_0^L - i a k_x v_x^{*} u_y \bigg|_0^L - ia k_x v_y^{*} u_x \bigg|_0^L - (a+b) v_y^{*} \left( \partial_y u_y \right)|_0^L + (a+b) \left( \partial_y v_y^{*}\right) u_y\bigg|_0^L \nonumber \\ 
    &\quad - i (c+d) k_z v_y^{*} u_z\bigg|_0^L - i(c+d)k_z v_z^{*} u_y\bigg|_0^L - d v_z^{*} \left( \partial_y u_z\right)|_0^L + d \left( \partial_y v_z^{*}\right) u_z \bigg|_0^L \nonumber \\ 
    &= -b v_x^{*}(L) u_x'(L) + b v_x^{*}(0) u_x'(0) + b v_x^{*'}(L) u_x(L) - ia k_x v_x^{*}(L) u_y(L) - i a k_x v_y^{*}(L) u_x(L) \nonumber \\ 
    &\quad - (a+b) v_y^{*}(L) u_y'(L) + (a+b) v_y^{*}(0) u_y'(0) + (a+b) v_y^{*'}(L) u_y(L) - i(c+d) k_z v_y^{*}(L) u_z(L) \nonumber \\ 
    &\quad - i(c+d) k_z v_z^{*}(L) u_y(L) - d v_z^{*}(L) u_z'(L) + d v_z^{*}(0) u_z'(0) + d v_z^{*'}(L) u_z(L) \nonumber  
\end{align}

where we used open boundary conditions ($\pmb{u}(0)=0$). We also use the free boundary conditions (Eqs(\ref{xzbcxy})-\ref{xzbcyy}) i.e.

\begin{align}
    & u_x'(L) + i k_x u_y(L) = 0 \nonumber \\
    & u_z'(L) + i k_z u_y(L) = 0 \nonumber \\
    & (a+b) u_y'(L) + (a-b) i k_x u_x(L) + c i k_z u_z(L) = 0 
\end{align}
We have
\begin{align}
    \chi &= ib k_x v_x^{*}(L) u_y(L) + b v_x^{*}(0) u_x'(0) + b v_x^{*'}(L) u_x(L) - ia k_x v_x^{*}(L) u_y(L) - i a k_x v_y^{*}(L) u_x(L) \nonumber \\ &+ i(a-b)k_x v_y^{*}(L) u_x(L) + i k_z c v_y^{*}(L) u_z(L) + (a+b) v_y^{*}(0) u_y'(0) + (a+b) v_y^{*'}(L) u_y(L) \nonumber \\ &- i(c+d) k_z v_y^{*}(L) u_z(L) - i(c+d) k_z v_z^{*}(L) u_y(L) + i d k_z v_z^{*}(L) u_y(L) + d v_z^{*}(0) u_z'(0) + d v_z^{*'}(L) u_z(L) \nonumber \\ 
    &= \Big[(a+b) v_y^{*'}(L) - i(a-b) k_x v_x^{*}(L) - i c k_z v_z^{*}(L) \Big] u_y(L) + b \Big[v_x^{*'}(L) - i k_x v_y^{*} (L) \Big] u_x(L) \nonumber \\ &+ d \Big[ v_z^{*'}(L) - i k_z v_y^{*}(L)\Big] u_z(L) + b v_x^{*}(0) u_x'(0) + (a+b) v_y^{*}(0) u_y'(0) + d v_z^{*}(0) u_z'(0) 
\end{align}

For $\chi=0$, we need $\pmb{v}$ to obey the same boundary conditions as $\pmb{u}$. Therefore, $D(H_0) = D(H_0^\dagger)$ and $H_0$ is self-adjoint. 

\subsubsection{Hermiticity of Eq(\ref{surfh1h2})}\label{Hermh1}

Consider $\langle \pmb{v}|H_1|\pmb{u} \rangle$ where $H_1$ is given by Eq(\ref{surfh1h2}). 

\begin{align}
    \langle \pmb{v}| H_1 | \pmb{u}\rangle &= - \langle \pmb{v}| \frac{\partial \theta}{\partial y} h_1 | \pmb{u}\rangle - \langle \pmb{v}| \frac{\partial^2 \theta}{\partial y^2} h_2 | \pmb{u}\rangle \nonumber \\
    &= - \frac{\partial \theta}{\partial y} \langle \pmb{v}|  h_1 | \pmb{u}\rangle - \langle \pmb{v}| \frac{\partial \theta}{\partial y} \partial_y \left( h_2 | \pmb{u}\rangle \right) \nonumber \\
    &= - \frac{\partial \theta}{\partial y} \langle \pmb{v}|  h_1 | \pmb{u}\rangle - \frac{\partial \theta}{\partial y} \langle \pmb{v}|  \partial_y \left( h_2 | \pmb{u}\rangle \right) 
\end{align}

From Eq(\ref{surfh1h2}), we see that $h_1^\dagger = h_1$ and $h_2^\dagger = -h_2$, which makes $h_1$ hermitian and $h_2$ anti-hermitian. 
The first term $- \frac{\partial \theta}{\partial y} \langle \pmb{v}|  h_1 | \pmb{u}\rangle $ is hermitian as $h_1^\dagger = h_1$. The second term $- \frac{\partial \theta}{\partial y} \langle \pmb{v}|  \partial_y \left( h_2 | \pmb{u}\rangle \right) $ is also hermitian because it consists of the product of two anti-hermitian operators $\partial_y$ and $h_2$.

\begin{figure*}
 \includegraphics[width=\textwidth]{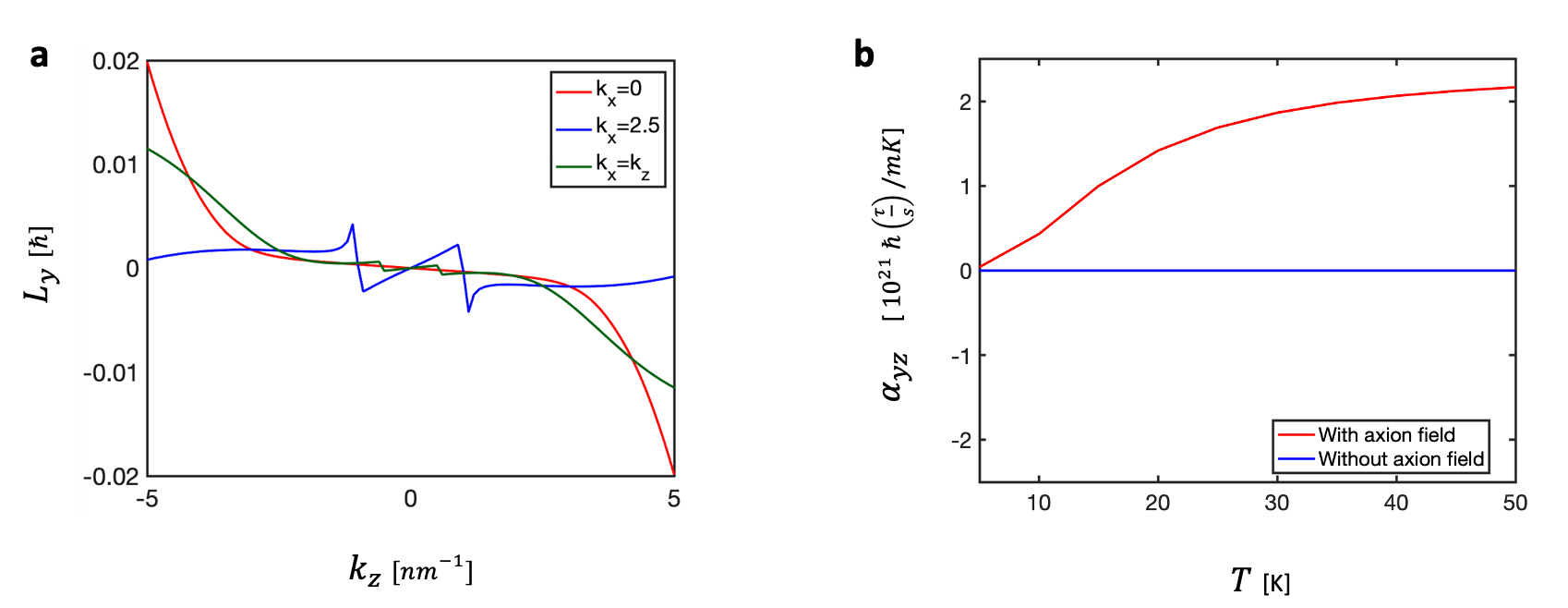}
 \caption{(a) The out of plane ($y$ direction) phonon angular momentum of the interface mode at $k_x=0$ (red) , $k_x=2.5 nm^{-1}$ (blue) and $k_x=k_z$ (dark green). (b) The $y$-direction angular momentum response to the $z$-direction thermal gradient,  $\alpha_{yz}$, of the interface phonon mode as a function of temperature in the presence (red) and absence (blue) of the axion field. We use $r=.8 $; $s=.713 $; $t=.9 $ in units of $10^{-12} m^4/s^2$. }
 \label{fig:S9}
 \end{figure*}

\section{Angular momentum of phonon modes}\label{appangmon}

\subsection{Derivation of total angular momentum for a lattice configuration}

In this section, we outline the angular momentum calculation for phonon modes. This formalism can be applied to both the interfacial phonon mode at the axion domain wall (presented in the main text) as well as the surface phonon mode described in Sec. \ref{secsurf}. Phonon angular momentum has been previously described in Refs. \cite{zhang2014angular,hamada2018phonon}. We consider a lattice model along the $y$ direction, where $k_x$ and $k_z$ are good quantum numbers, but $k_y$ is not a good quantum number due to the domain wall structure. We write the mode expansion for the displacement field and the corresponding canonical momentum in terms of the $a$ and $a^\dagger$ as

\begin{align}
    u_i (x,y,z) &= \int \frac{dk_x }{(2\pi)}\frac{dk_z }{(2\pi)} \sum_{\alpha,n} \sqrt{\frac{\hbar}{2 \rho \omega^\alpha_n (k_x, k_z)}} \epsilon^\alpha_i (k_x, k_z,y,n)  \left[ a^\alpha_n (k_x, k_z) e^{i \left( k_x x + k_z z - \omega^\alpha_n (k_x,k_z) t \right)} \right] + h.c \\
    \Pi_i (x,y,z) &= -i \int \frac{dk_x }{(2\pi)}\frac{dk_z }{(2\pi)} \sum_{\alpha,n} \sqrt{\frac{\hbar \rho \omega^\alpha_n (k_x, k_z)}{2}}  \epsilon^\alpha_i (k_x, k_z, y, n)  \left[ a^\alpha_n (k_x, k_z) e^{i \left( k_x x + k_z z - \omega^\alpha_n (k_x,k_z) t \right)} \right] + h.c
\end{align}
where the index  $\alpha$ in $\epsilon^\alpha$ labels ${ \alpha =1,2,3}$ which represent 3 possible polarizations. In the isotropic limit, these correspond to 1 longitudinal polarization and 2 transverse polarizations.  $n$ labels the quantum number corresponding to the $y$-direction confinement, i.e. $n=1,2...,N+1$, for $N+1$ quantum well states. The operator $a^\alpha_n (k_x, k_z)$ 
creates a phonon described by the eigenvector $\Vec{\epsilon}^\alpha (k_x, k_z, y, n)$ with polarization $\alpha$, the quantum number $n$, momenta $k_x$ and $k_z$. $ \omega^\alpha_n$ labels the eigen frequency. We now define $\Vec{r} = x \hat{e}_x + z \hat{e}_z$ and $\Vec{k} = k_x \hat{e}_x + k_z \hat{e}_z$. The angular momentum $j_i(x,y,z)$ is given by

\begin{align} \label{smallj}
    &j_i (x,y,z) =  \epsilon_{ijl} u_j(x,y,z) \Pi_l (x,y,z) \nonumber \\
    &=  - i \epsilon_{ijl} \frac{\hbar}{2} \sum_{\alpha, \beta, n, m} \int \frac{d^2k}{(2 \pi)^2} \frac{d^2k'}{(2 \pi)^2} \sqrt{\frac{\omega^\beta_m(\Vec{k})}{\omega^\alpha_n (\Vec{k'})}}\left[ \epsilon_j^\alpha (\Vec{k}, y, n) a^\alpha_n (\Vec{k}) e^{i\left(\Vec{k}\cdot\Vec{r} - \omega^\alpha_n (\Vec{k})t\right)} 
    + \epsilon_j^\alpha (\Vec{k}, y, n)^* a^\alpha_n(\Vec{k}) ^\dagger e^{-i\left(\Vec{k}\cdot\Vec{r} - \omega^\alpha_n (\Vec{k})t\right)} \right] \nonumber \\ 
    &\times \left[ \epsilon_j^\beta (\Vec{k'}, y, m) a^\beta_m (\Vec{k'}) e^{i\left(\Vec{k'}\cdot\Vec{r} - \omega^\beta_m (\Vec{k'})t\right)} -  \epsilon_j^\beta (\Vec{k'}, y, m)^* a^\beta_m(\Vec{k'}) ^\dagger e^{-i\left(\Vec{k'}\cdot\Vec{r} - \omega^\beta_m (\Vec{k'})t\right)} \right].
\end{align}

The total angular momentum is defined as $J_i = \sum_y \int d^2r j_i (x,y,z)$. We integrate Eq(\ref{smallj}) over the $xz$ surface and sum over all $y$ sites. We make use of the relation $\int d^2 r e^{i \left( \Vec{k} - \Vec{k}\right)\cdot \Vec{r} } = (2\pi)^2 \delta^2(\Vec{k}-\Vec{k'})$ and find

\begin{align} \label{smalljM}
    J_i &= - i \epsilon_{ijl} \frac{\hbar}{2}  \sum_{\alpha,\beta,n, m, y} \int \frac{d^2k}{(2\pi)^2} \sqrt{\frac{\omega^\beta_m(\Vec{k})}{\omega^\alpha_n (\Vec{k})}} \Big[ \epsilon_j ^\alpha (\Vec{k},y, n)^* \epsilon_l^\beta (\Vec{k},y, m) a^\alpha_n (\Vec{k})^\dagger a^\beta_m (\Vec{k}) e^{i\left( \omega^\alpha_n (\Vec{k}) - \omega^\beta_m (\Vec{k})\right) t} \nonumber \\ 
    & - \epsilon_j ^\alpha (\Vec{k},y, n) \epsilon_l^\beta (\Vec{k},y, m)^* a^\alpha_n (\Vec{k}) a^\beta_m (\Vec{k})^\dagger e^{-i\left( \omega^\alpha_n (\Vec{k}) - \omega^\beta_m (\Vec{k})\right) t}\Big].  
\end{align}

Since the system is in equilibrium, we have neglected terms of the form $a a$ and $a^\dagger a^\dagger$. Furthermore, for the second term on the right side of Eq(\ref{smalljM}), we can switch the indices $j$ and $l$ and write $-\epsilon_{ijl} \epsilon_j^\alpha(n) (\epsilon_l^\beta(m)) ^* = \epsilon_{ijl} \epsilon_l^\alpha(n)  \epsilon_j ^\beta(m) ^*$. We define $\left( M_i\right)_{jl} = -i \hbar \epsilon_{ijl}$ and find

\begin{align}
    J_i &= \frac{1}{2}  \sum_{\alpha,\beta,m,n,y} \int \frac{d^2k}{(2\pi)^2} \sqrt{\frac{\omega^\beta_m(\Vec{k})}{\omega^\alpha_n (\Vec{k})}} \epsilon_j ^\alpha (\Vec{k},y,n)^* \left( M_i\right)_{jl} \epsilon_l^\beta (\Vec{k},y,m) \Big[  a^\alpha_n (\Vec{k})^\dagger a^\beta_m (\Vec{k}) e^{i\left( \omega^\alpha_n (\Vec{k}) - \omega^\beta_m (\Vec{k})\right) t} \nonumber \\ 
    & + a^\alpha_n (\Vec{k}) a^\beta_m (\Vec{k})^\dagger e^{-i\left( \omega^\alpha_n (\Vec{k}) - \omega^\beta_m (\Vec{k})\right) t}\Big].
\end{align}

We use the commutation relation $\left[ a^\alpha_n (\Vec{k}), a^\beta_m (\Vec{k'})^\dagger \right] = (2 \pi)^2 \delta^2\left(\Vec{k} - \Vec{k'} \right) \delta_{\alpha,\beta} \delta_{n,m} $. In equilibrium, $\langle  a^\alpha_n (\Vec{k}) ^\dagger a^\beta_m (\Vec{k})  \rangle_0 = f\left( \omega^\alpha_n (\Vec{k})\right) \delta_{\alpha, \beta} \delta_{n,m}$, where $f\left( \omega^\alpha_n (\Vec{k})\right)$ is the Bose distribution. Therefore, the total angular momentum is given by

\begin{equation} \label{angmomtot}
    J_i = \int \frac{d^2k}{(2\pi)^2} \sum_{\alpha,n, y} l^\alpha_i (\Vec{k},y,n) \left[ f\left( \omega^\alpha_n (\Vec{k})\right) + 1/2 \right]  , l^\alpha_i (\Vec{k},y,n) = \epsilon^\alpha_j(\Vec{k},y,n)^* \left(M_i\right)_{jl} \epsilon^\alpha_l(\Vec{k},y,n).
\end{equation}

\subsection{Effect of thermal gradient}

We now consider the system away from equilibrium. In order to invoke Boltzman transport theory \cite{girvin2019modern,hamada2018phonon}, we assume small deviation away from equilibrium. We also assume that the phonon relaxation process quickly brings the system back to thermal equilibrium, which can be characterized by the relaxation time $\tau$.  In this semiclassical picture, both $\Vec{k}$ and $\Vec{r}$ are well-defined and we conisder the phase space $(x,z,k_x,k_z)$. The distribution function $f^{\alpha}(\Vec{r},\Vec{k}, n, t)$ is the probability of an electron at $\Vec{r}$ occupying the state $\Vec{k}$ with $n$ and $\alpha$ at time $t$. The equilibrium state of phonons satisfies the Bose-Einstein distribution. In the absence of collisions, we have for a particular $n$ and $\alpha$,

\begin{align}
    \Vec{r}(t-dt) &= \Vec{r}(t) - \Vec{v}^{\alpha}_n(\Vec{k})dt, \\
    \Vec{k}(t-dt) &= \Vec{k}(t) - \frac{1}{\hbar} \Vec{F}(\Vec{r},\Vec{k},t) dt,
\end{align}
where  $\Vec{v}^{\alpha}_n(\Vec{k})$ is the group velocity given by $\Vec{v}^{\alpha}_n(\Vec{k}) = \frac{1}{\hbar} \frac{\partial E^\alpha_n(\Vec{k})}{\partial \Vec{k}} = \frac{1}{\hbar} \frac{\partial }{\partial \Vec{k}}  \left(\rho \omega^\alpha_n (\Vec{k})^2\right)$. For the next few lines of the derivation, we omit the indices $n$ and $\alpha$ but bring them back at the end of the derivation. Liouville's theorem requires 
\begin{align} 
f(\Vec{r},\Vec{k},t) &= f\left( \Vec{r} - \Vec{v}(\Vec{k})dt, y, \Vec{k} -dt \right) \nonumber \\
&= f(\Vec{r}, \Vec{k},t) - \left( \Vec{\nabla}_r f\right)\cdot \Vec{v}(\Vec{k}) dt - \frac{\partial f}{\partial t} dt   
\end{align}

Rearranging the terms, we have

\begin{equation}
    \frac{\partial f}{\partial t} + \Vec{v} \cdot \left(\Vec{\nabla}_r f \right)  = 0   
\end{equation}

We add the collision term $\left( \frac{\partial f}{\partial t} \right)_{\text{coll}}$ to the right-hand side of the above equation to take into account the relaxation process. In the relaxation-time approximation, the phenomenological ansatz is \begin{equation}
    \left( \frac{\partial f}{\partial t} \right)_{\text{coll}} = - \frac{1}{\tau} \left(  f(\Vec{r},\Vec{k},t) - f^0_{\Vec{k}}\right), 
\end{equation}  
where $\tau$ is the relaxation time and $f^0_{\Vec{k}}$ is the equilibrium distribution. Therefore, we have 

\begin{equation}
    \Vec{v}(\Vec{k})\cdot \Vec{\nabla}_r f = \left( \frac{\partial f}{\partial t} \right)_{\text{coll}} = - \frac{1}{\tau} \left(  f(\Vec{r},\Vec{k},t) - f^0_{\Vec{k}}\right)
\end{equation}

We restore the indices $n$ and $\alpha$, so

\begin{align}
    f\left( \omega^\alpha_n (\Vec{k}) \right) &= f^0 \left( \omega^\alpha_n (\Vec{k})\right) - \tau \Vec{v}^\alpha_n(\Vec{k}) \cdot \Vec{\nabla} _r f^0 \left( \omega^\alpha_n(\Vec{k})\right) \nonumber \\
    &= f^0 \left( \omega^\alpha_n (\Vec{k})\right) - \tau v_{i,n}^\alpha(\Vec{k})  \frac{\partial f^0}{\partial T} \frac{\partial T}{\partial x_i}. \label{noneqf}
\end{align}

Plugging Eq(\ref{noneqf}) into Eq(\ref{angmomtot}), we have

\begin{equation} \label{angmomgrad}
    J_i = \int \frac{d^2k}{(2\pi)^2} \sum_{\alpha,y,n} l_i^\alpha (\Vec{k},y,n) \left[ f\left( \omega^\alpha_n (\Vec{k})\right) + 1/2 \right] = - \tau  \int \frac{d^2k}{(2\pi)^2} \sum_{\alpha,y,n} l^\alpha_i (\Vec{k},y,n) v^{\alpha}_{j,n} (\Vec{k})  \frac{\partial f^0}{\partial T} \frac{\partial T}{\partial x_j},
\end{equation}
where we have used the fact that for a time reversal invariant system, $l^\alpha(-\Vec{k},y, n) = - l^\alpha(\Vec{k},y, n)$ but $f^0\left(\omega^\alpha_n (-\Vec{k}) \right) = f^0\left(\omega^\alpha_n (\Vec{k}) \right)$ i.e. 

\begin{align}
    &\int \frac{d^2k}{(2\pi)^2}  l_i^\alpha(\Vec{k},y,n) = 0 \\
    &\int \frac{d^2k}{(2\pi)^2}  l_i^\alpha(\Vec{k},y,n) f^0\left(\omega^\alpha_n (\Vec{k}) \right) =0  
\end{align}

We rewrite Eq(\ref{angmomgrad}) as $J_i = \alpha_{ij} \frac{\partial T}{\partial x_j} $ and the response tensor is given by

\begin{equation}
        \alpha_{ij} = - \tau  \int \frac{d^2k}{(2\pi)^2} \sum_{\alpha,y,n} l^\alpha_i (\Vec{k},y,n) v^{\alpha}_{j,n}(\Vec{k})  \frac{\partial f^0}{\partial T} = - \frac{\tau}{A_{xz}} \sum_{\Vec{k},\alpha,y,n} l^\alpha_i (\Vec{k},y,n) v^{\alpha}_{j,n}(\Vec{k})  \frac{\partial f^0}{\partial T} 
\end{equation}

\subsection{Symmetry properties of the response tensor $\alpha_{ij}$}

In order to probe the angular momentum response of the interfacial or surface phonon modes, the thermal gradient must be applied along the plane perpendicular to the axion domain wall i.e. along the $xz$ plane. The response function forms a $3\times 2$ matrix as the temperature gradient is restricted to the film plane. In the absence of the axion term, the symmetries of the $xz$ plane are generated by $m_x: x \rightarrow -x$ and $m_z: z \rightarrow -z$. The axial tensor must obey $m_x \alpha m_x^{-1} = - \alpha$ and $m_z \alpha m_z^{-1} = - \alpha$. So

\begin{equation}
    \alpha = \begin{pmatrix} 
     0  & \alpha_{xz}\\
     0  & 0 \\
     \alpha_{zx}  & 0
    \end{pmatrix}
\end{equation}

The axion surface term $F_1$ breaks $m_x$, therefore, $\alpha_{zy}, \alpha_{yz} \neq 0$. In the presence of the axion term 

\begin{equation}
    \alpha = \begin{pmatrix} 
     0 &  \alpha_{xz}\\
     0 & 
     \alpha_{yz}\\
     \alpha_{zx}  & 0
    \end{pmatrix},
\end{equation}
which provides an additional $\alpha_{yz}$ term due to strain (axion term is created by strain) corresponding to the $y$-direction phonon angular momentum response due to the $z$-direction thermal gradient of the interfacial phonon mode. The phonon angular momentum in the momentum space is shown in Fig. \ref{fig:S9}a and the thermal gradient induced phonon angular momentum as a function of temperature is shown in Fig. \ref{fig:S9}b, in which the axion term is found to be essential for the component $\alpha_{yz}$. $\alpha_{yz}$ is calculated to be $\sim 10^{21} \hbar \left(\frac{\tau}{s} \right) m^{-1} K^{-1} \sim 10^{-13} \left(\frac{\tau}{s} \right) Jsm^{-1}K^{-1} $. It should be noted that the total angular momentum defined in Eq.(\ref{angmomtot}) is for a two dimensional slab, instead of a three dimensional (3D) bulk, so the unit of the coefficient $\alpha_{yz}$ is different from that defined in Ref.\cite{hamada2018phonon, liu2022probing}. For a comparison with the value of $\alpha$ for 3D bulk, we choose the decaying length $l_d$ of the interfacial modes around $10\sim 100$nm as the inverse of decaying length  $\lambda=1/l_d$ is around $0.01\sim 0.1 nm^{-1}$. Using this length scale for the y direction, we can estimate $\alpha_{yz}/l_d \approx 10^{-6} \sim 10^{-5} \left(\frac{\tau}{s} \right) Js m^{-2} K^{-1}$, which is comparable to the values of $\alpha \sim 10^{-6} \left(\frac{\tau}{s} \right) Js m^{-2} K^{-1}$ from the previous studies of temperature gradient induced phonon angular momentum in bulk materials \cite{hamada2018phonon, liu2022probing}.

\bibliographystyle{apsrev4-2}

\bibliography{ref}

%\end{document}

\end{document}